\documentclass[useAMS,usenatbib,babel,times,superscriptaddress]{mnras}

\usepackage[english,english]{babel}
\usepackage{amsmath}
\usepackage{amssymb,amsfonts,textcomp}
\usepackage{array}
\usepackage{supertabular}
\usepackage{hhline}
\usepackage{hyperref}
\usepackage[usenames]{color}
\usepackage{algpseudocode}
\algrenewcommand\textproc{}% Disable uppercase in functions

\def\gtrsim{\lower.5ex\hbox{$\; \buildrel > \over \sim \;$}}
\usepackage{graphicx}

\def \apjl {{\it ApJ Let.}}
\def \apjs {{\it ApJ Sup.}}

\def \aap {{\it A\&A}}

\DeclareGraphicsExtensions{.jpg,.mps,.pdf,.png}

\newcommand{\mR}{{\cal{R}}}
\newcommand{\mL}{{\cal{L}}}
\newcommand{\mA}{{\cal A}}
\newcommand{\mB}{{\cal B}}

\newcommand{\mD}{{\cal D}}
\newcommand{\mP}{{\cal P}}
\newcommand{\mS}{{\cal S}}
\newcommand{\mF}{{\cal{F}}}

\newcommand{\vx}{\textbf{x}}
\newcommand{\vecr}{\textbf{r}}

\newcommand{\vk}{\textbf{k}}

\newcommand{\dd}{\hbox{d}}
\newcommand{\ii}{\hbox{i}}

\newcommand{\hrho}{{\hat \rho}}
\newcommand{\hs}{{\hat{s}}}

\newcommand{\lin}{{\rm lin}}

\def\red{}

\begin{document}

\title[Correlations of multi-cell densities]{
The large-scale correlations of multi-cell densities and profiles: implications for cosmic variance estimates}
\author[ S.~Codis,  F.~Bernardeau,  and C.~Pichon]{
Sandrine Codis$^{1,2}$\thanks{codis@cita.utoronto.ca}, Francis Bernardeau$^{2,3}$,   Christophe Pichon$^{2,4}$ 
\vspace*{6pt}\\
\noindent 
$^{1}$ Canadian Institute for Theoretical Astrophysics, University of Toronto, 60 St. George Street, Toronto, ON M5S 3H8, Canada\\
$^{2}$ 
Sorbonne Universit\'es, UPMC Univ Paris 6 \& CNRS, UMR 7095, Institut d'Astrophysique de Paris,
 98 bis bd Arago, 75014 Paris, France\\
$^{3}$ CNRS \& CEA, UMR 3681, Institut de Physique Th\'eorique, F-91191 Gif-sur-Yvette, France\\
$^{4}$ Korea Institute of Advanced Studies (KIAS) 85 Hoegiro, Dongdaemun-gu, Seoul, 02455, Republic of Korea
}
\maketitle
\begin{abstract}
{In order to quantify the error budget in the measured probability distribution functions of cell densities, the two-point statistics of cosmic densities in concentric spheres 
is investigated.
Bias functions are introduced as the 
ratio of their two-point correlation function to the  two-point correlation of the underlying dark matter distribution.
They  describe how cell densities  are spatially correlated. They are computed here via the so-called large deviation principle in the quasi-linear regime.  Their  \textit{large-separation limit} is presented    and 
successfully compared to simulations  for density and density slopes:
this regime is shown to be rapidly reached allowing to get sub-percent precision  for a wide range of densities and variances.
 The corresponding asymptotic limit provides an estimate of the cosmic variance of standard concentric cell statistics applied to finite surveys.
More generally, no assumption on the separation is required for some specific moments of the two-point statistics,  for instance when  predicting the generating function of cumulants containing any powers of concentric densities in one location and one power of density at some \textit{arbitrary} distance from the rest. This exact ``one external leg'' cumulant generating function is used in particular to probe the rate of convergence of the large-separation approximation.
}
\end{abstract}

%%%%%%%%%%
\section{Introduction}

The geometry of the large-scale structure of the Universe  puts very tight constraints on  cosmological models. Deep spectroscopic surveys, like Euclid \citep{Euclid} or DESI \citep{DESI}, will soon allow us to study the details of structure formation at different epochs with unrivaled precision and therefore offer insight into the engine of cosmic acceleration. 
In order to reach percent precision on the equation of state of dark energy, astronomers are facing various challenges: non-linear gravitational evolution \citep{2002PhR...367....1B}, redshift space distortions \citep{Kaiser87,tns}, bias \citep{Kaiser84,dekel87}, intrinsic alignments \citep{2015SSRv..193...67K}, baryonic physics \citep{2015JCAP...12..049S} to name a few. Probing the non-linear regime increases the number of   modes  used
to better constrain cosmological parameters. Hence, theorists need to investigate alternatives to the standard N-point correlation functions \citep{Scoccimarro98} that Perturbation Theory can only predict in the weakly non-linear regime \citep{2015arXiv151004075L}. They must find new observables that can be predicted from first principles, and do not rely solely  on  very large  simulations of the Universe produced with  hundreds of millions of CPU hours. 

It has been argued \citep{Bernardeau14,Bernardeau15,uhlemann16} that the statistics of cosmic densities in concentric spheres can leverage cosmic parameters 
competitively, as the corresponding spherical symmetry allows for analytical predictions in the mildly non-linear regime, beyond 
what is commonly achievable via other statistics such as correlation functions. Indeed, the zero variance limit  of the cumulant generating functions
yields estimates of the joint probability distribution function (PDF hereafter) which seems to match simulations in the regime of variances of order unity {\red as shown by \cite{2002A&A...382..412V,Bernardeau14,Bernardeau15} building upon some earlier investigations by  \cite{1989A&A...220....1B,1992ApJ...390L..61B}}. This success has been recently shown to correspond to a large-deviation principle in the context of cosmic structure formation \citep{LDPinLSS} which is based on the sole assumption that the variance of the field is small. {\red The corresponding predictions for the successive reduced cumulants were shown to be in excellent agreement with simulations for scale above a few Mpc$/ h$ (e.g \cite{1995MNRAS.274.1049B})}. It has to be contrasted with the commonly used polyspectra predicted by Perturbation Theory which typically require the fluctuations of the field to be small everywhere. For instance, unless large-deviation predictions, the Edgeworth expansion typically breaks down for $|\delta|>\sigma$.
  
Hence it is of interest to quantify the error budget for such large-deviation estimators, while accounting for the expected long-range clustering
 within realistic surveys of finite extent. Indeed, in practice, measurements of cosmic densities  cannot be carried out for different realisations i.e different universes but in one finite part of our Universe as mapped by  surveys like the SDSS \citep{SDSS} or DES \citep{DES}. The density in spheres  drawn from those surveys are not independent. Consequently, this dependence  induces errors which, at first order, are dominated by the two-point correlation between spheres, as shown by \cite{1987ApJ...320...13M,1995ApJS...96..401C,1996ApJ...470..131S,2011ApJ...731..113M}.
   Estimating the two-point correlations of concentric cosmic densities \citep{1996A&A...312...11B,1999MNRAS.310..428S} is therefore important  to mitigate the cosmic variance on the measurement of  their one-point statistics.   Once a model for these correlations exists, it  can be integrated into maximum likelihood estimators for the underlying cosmic parameters.
   
 Besides error statistics, the study of this  two-point clustering statistics of concentric spheres is also interesting in its own right,  as it allows one to investigate how the densest regions of space -- where dark halos usually reside -- are clustered in the quasi-linear regime, 
 which in turn  sheds light on  the so-called biasing  between dark matter and halos:
 {\red a common, apparently good assumption \citep{2011MNRAS.413.1961L}, is that haloes
}
correspond to peaks of the density field, they  are therefore not  a  fair tracer of that field. For Gaussian random fields,
\cite{Kaiser84} showed that 
in the high contrast $\nu$, high separation limit, the correlation function, $\xi_{>\nu}$, between two regions lying above a threshold $\nu$ reads
\begin{equation}
\xi_{>\nu}\approx \nu^{2}\xi, 
\label{eq:xikaiser}
\end{equation}
so that the correlation function of high-density regions decreases more slowly than the density field correlation function, $\xi$, with an amplification factor or clustering bias that is proportional to the threshold squared. This analysis can also be restricted to the peaks of the density field above a given threshold following the seminal paper by \cite{BBKS}.

Building upon the idea  that the density of large clusters must be strongly clustered compared to the density of galaxies,  
\cite{davis85} popularized the notion of linear bias between the galaxy distribution and dark matter density field
\begin{equation}
\delta_\mathrm{g}=b_{1}\delta_{\rm DM}\,,
\end{equation}
where $b_{1}$ is assumed to be a constant which was shown to be a good approximation on sufficiently large scales \citep{2011MNRAS.415..383M}.
However, given the complexity of galaxy formation, the validity of this approximation is likely to be somewhat narrow. 
\cite{1993ApJ...413..447F} proposed to extend the linear bias approach and to introduce the so-called local bias model in which the full Taylor expansion of  the relation $\delta_{g}={\cal F}(\delta_{\rm DM})$ is considered. This parametrization can be studied  from a Eulerian or a Lagrangian point of view, meaning at final or initial time,  \citep{2000MNRAS.318L..39C} and can account for stochasticity (\cite{dekel99}), time-evolution (\cite{nusser94,fry96,teg98}), non-linearity (\cite{pen98,guo09}), scale-dependence (\cite{lumsden89,2010PhRvD..81f3530G}) or non-locality in time \citep{2015JCAP...11..007S}. Non-local effects can also be addressed by parametrizing the dependence of the galaxy density with other operators such as the tidal tensor or velocity shear as long as they preserve the symmetry and equivalence \citep{2009JCAP...08..020M,2012PhRvD..86h3540B,2015JCAP...07..030M}.
 De facto, this bias is very likely to depend on the population of galaxies and to be not only a function of the density but also temperature, merging history and other galaxy properties, introducing some scatter  in the galaxy-matter density relation. 
Quantifying bias is crucial in cosmology if one wants to measure the cosmological parameters encoded in  $\xi$  from, e.g. the  correlation function of galaxies. It is also an interesting quantity to measure in its own right, as it carries information on the physics of galaxy formation. 
Numerical simulations are  very efficient  at accurately predicting and calibrating galactic observables. Improvements of our understanding of galaxy formation and its implementation in simulations is a timely and fast moving topic of research.

However, it has to be noted that the scale-dependence of the {\red halo bias (defined as the ratio of the halo-halo to dark matter correlation functions)} is strong on small scales but remains weak with typical deviations of less than a few percents on scales above 20 Mpc/h \citep{2015MNRAS.453.1513C}. 
It would therefore be interesting to predict {\red the bias of specific regions (for instance high density and/or negative slope as a proxy for dark halos)} from first principles not only in the linear stage of structure formation but also in the subsequent mildly non-linear regime.

The aim of this paper is to quantify the effect of cosmic variance  on count-in-cells statistics in the mildly non-linear regime. Specifically we will estimate
 the  two-point correlations of concentric cosmic densities at different positions in the field in the (not so) large separation limit,
 while relying on 
the symmetric framework of the spherical collapse, which leads to surprisingly accurate predictions. {\red It will allow us in particular to define the bias associated with concentric cosmic densities and provide accurate predictions up to a few Mpc$/h$.}

This paper is organized as follows. Section~\ref{sec:jointstat} 
 predicts the bias and error budget expected in standard concentric count-in-cells. Section~\ref{sec:cumulant} shows how to compute the generating function of the cumulants containing any power of the densities in one location and one power of the density at an arbitrary distance. Section~\ref{sec:bias} 
 predicts the bias functions in the large separation limit while section~\ref{sec:prediction} applies the formalism to realistic power spectra.
Section~\ref{sec:validation} validates this large-separation and large-deviation approximation on simulations {\red before Section~\ref{sec:illustration} illustrates how bias functions can be used to predict the expected error budget when estimating the PDF of the cosmic density in concentric spheres}. Finally, Section~\ref{sec:conclu} wraps up. Appendices~\ref{decimation}, \ref{pkbgarg}, \ref{discretecounts}, \ref{sec:consistency}, \ref{sec:largedensity} and \ref{sec:approx-slope} respectively present the radii decimation used in the main text, an argument in favour of the large-distance  factorizations, some calculations of cosmic variance for discrete counts, a consistency check of the measurements of the bias function by two distinct methods and analytical asymptotes for the density and slope bias.

%%%%%%%%%%
\section{Statistics of sets of  count-in-cells }
\label{sec:jointstat}

In a survey or a simulation,  measurements of concentric densities are carried as follows: sets of concentric cells are drawn randomly 
or regularly across the field at some finite distance to each other. In cosmology, this field has long-range correlations which will break the assumption that 
each set of concentric spheres can be considered independently.
 In order to estimate the induced bias and variance, let us study the joint statistics of these sets.

\subsection{The bias function 
of concentric spheres}

\begin{figure}
\includegraphics[width=8cm]{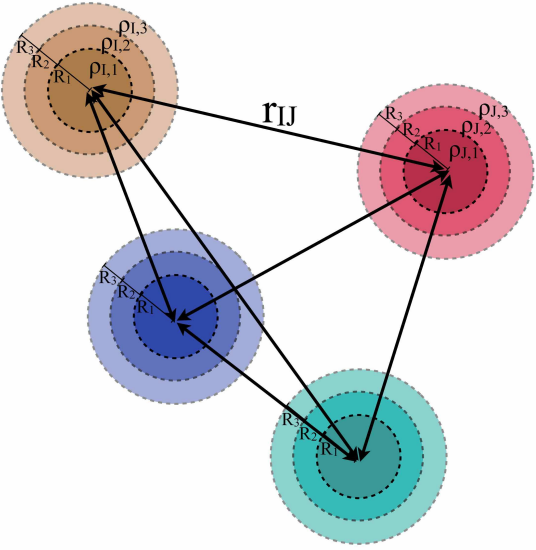}
   \caption{The configuration of  spherical cells considered in this paper
which is made of  multiple sets  of concentric spheres separated by 
distances $r_{\rm{IJ}}$. Their respective density,  $\rho_{\rm{I},i}$, corresponds to a set of $n$ spheres of same radii $R_{\rm I,i}\equiv R_{i}$.
   \label{fig:config}}
\end{figure}
Let us  consider multiple sets (labelled from $\rm I=1$ to $\rm N_{t}$)  of $n$ concentric spheres (labelled from $k=1$ to $n$) of radii $R_{\rm I,k}\equiv R_{k}$ separated by 
distances $r_{\rm{IJ}}$, 
and define the corresponding measured 
%(hence the hat) 
densities $\{\rho_{\rm I,k}\}$ (see Fig.~\ref{fig:config} for an illustration).
The joint PDF of those $\rm N_{t}$ sets, 
\begin{equation}
\mP(\{\rho_{1,k}\},\dots,\{\rho_{{\rm N_{t}},k}\};\{r_{\rm{IJ}}\}) \,, \label{eq:defPDF}
\end{equation}
{\red determines the full hierarchy of cumulants} of the 
form
\begin{equation}
\langle 
\rho_{1,1}^{p_{1}}\dots\rho_{1,n}^{p_{n}}
{\rho_{2,1}^{q_{1}}}\dots{\rho}_{2,n}^{q_{n}} \dots 
{\rho_{\rm{{N_{t}}},1}}^{s_{1}}\dots{\rho}_{{\rm{N_{t}}},n}^{s_{n}}
\rangle_{c}\,. \label{eq:cummixed0}
\end{equation}
The purpose of this paper is to estimate the joint PDF $\mP(\{\rho_{1,k}\},\dots,\{\rho_{\rm{N_{t}},k}\};\{r_{\rm{IJ}}\})) $ in the large-separation limit,
where $r_{\rm{IJ}}\gg R_{\mathrm {max}}=\max_{j}R_j$.
In this limit, we will  demonstrate in Section~\ref{sec:np} that this PDF reads
\begin{align}
\mP(&\{\rho_{1,k}\},\dots,\{\rho_{{\rm N_{t}},k}\};
\{
r_{\rm{IJ}}
\}\gg R_j) 
=\nonumber\\
&\prod_{\rm I=1}^{\rm{N_{t}}} \mP(\{\rho_{{\rm I},k}\})
\left[1+\
\sum_{\rm{I<J}}
b(\{\rho_{{\rm I},k}\}) b(\{\rho_{{\rm J},k}\}) \xi(r_{\rm{IJ}})
\right]\,, \label{eq:fullPDFlargeseparation}
\end{align}
where $\Pi_{\rm I} \mP(\{\rho_{{\rm I},k}\})$ is the product of one-point PDFs,
 $\xi(r)$ is the underlying dark matter correlation function, and $b(\{\rho_{{\rm I},k}\})$ is some local bias function 
for the set I of $n$ concentric spheres. This is the count-in-cell analog of the so-called 
peak-background-split or clustering bias. 
Equation~(\ref{eq:fullPDFlargeseparation}) is the key result of this paper and will be used in 
the following sections to compute $b(\{\rho_{{\rm I},k}\})$ whose final expression is given by 
equations~(\ref{eq:defbias}) and (\ref{eq:defbiasPDF}) below. We will also show in Section~\ref{sec:properties} that the bias obeys $\int b(\rho){\cal P}(\rho) \dd \rho=0$ and $\int \rho\, b(\rho){\cal P}(\rho) \dd \rho=1$ so that the N-point PDF given in equation~(\ref{eq:fullPDFlargeseparation}) is normalised and its marginal in one location is exactly given by the one-point PDF.

Equation~(\ref{eq:fullPDFlargeseparation}) allows us to define the excess probability of having the  sets of densities
  $\{\rho_{{1},k}\},\dots,\{\rho_{{\rm N_{t}},k}\}$ separated by $\left\{r_\mathrm{IJ}\right\}$ as 
\begin{equation}
\xi_{\rm N_{t}}(\{\rho_{1,k}\},\dots,\{\rho_{{\rm N_{t}},k}\})=\sum_\mathrm{I<J}
b(\{\rho_{{\rm I},k}\}) b(\{\rho_{{\rm J},k}\})\xi(r_{\rm IJ}). \label{eq:fullPDFlargeseparation2}
\end{equation}
From equation~(\ref{eq:fullPDFlargeseparation2}), we see that the error in assuming that the draws of concentric densities
in simulations
are independent scales like the dark matter correlation\footnote{
In analogy with the corresponding situation for peaks, we can anticipate corrections involving derivative
of the dark matter correlation at shorter separations.}.

\subsection{The bias and variance  of concentric cumulants}

Let us now define the arithmetic mean over sets of concentric spheres 
 as 
\begin{equation}
\overline{\rho_{1}^{p_{1}}\dots\rho_{n}^{p_{n}}}\equiv \frac{1}{\rm N_{t}}\sum_{\mathrm I} \rho_{\mathrm I,1}^{p_{1}}\dots\rho_{\mathrm I,n}^{p_{n}}\,.\label{eq:Defestimator}
\end{equation}
This quantity naturally corresponds to what astronomers would measure in practice (spatial averages rather than ensemble averages).
Our purpose is to quantify the bias and the expected cosmic variance of this estimator.
Given equation~(\ref{eq:fullPDFlargeseparation}), one can check that the expectation of the arithmetic estimator 
defined by equation~(\ref{eq:Defestimator}) obeys
\begin{equation}
\langle \overline{\rho_{1}^{p_{1}}\dots\rho_{n}^{p_{n}}} \rangle_c =\langle {\rho_{1}^{p_{1}}\dots\rho_{n}^{p_{n}}} \rangle_c \,,
\end{equation}
so that the mean of the estimator given by
equation~(\ref{eq:Defestimator}) is unbiased at large distances.

%%%%%%%%%%%%%
Let us now estimate the cross correlation of this estimator,
$C_{\mathbf p \mathbf q}\equiv
\langle \overline{\rho_{1}^{p_{1}}\dots\rho_{n}^{p_{n}}}   \,\overline{\rho_{1}^{q_{1}}\dots\rho_{n}^{q_{n}}}
\rangle_c$
and express it in terms of moments of the bias function
\begin{multline}
C_{\mathbf p \mathbf q}=
\frac 1 {\mathrm{N_{t}}} \left\langle \rho_{1}^{p_{1}+q_{1}}\dots\rho_{n}^{p_{n}+q_{n}}\right\rangle_{c}
+
\frac 1 {\mathrm{N_{t}}^{2}}\sum_{\rm I\neq J}\xi(r_\mathrm{IJ})\times\\
%\partial_{\lambda_{1}\dots\lambda_{n}}^{p_{1}\dots p_{n}}b_{\varphi}(\{\lambda_{k}\})\Big|_{\lambda_{i}=0}\,
\left\langle b(\rho_{1}\!\dots\!\rho_{n})\rho_{1}^{p_{1}}\!\dots\!\rho_{n}^{p_{n}}\right\rangle_{c}
%\partial_{\lambda_{1}\dots\lambda_{n}}^{q_{1}\dots q_{n}}b_{\varphi}(\{\lambda_{k}\})\Big|_{\lambda_{i}=0}
\left\langle b(\rho_{1}\!\dots\!\rho_{n})\rho_{1}^{q_{1}}\!\dots\!\rho_{n}^{q_{n}}\right\rangle_{c}
\,.
\label{eq:Cpp}
\end{multline}
%where $\partial_{\lambda_{1}\!\dots\!\lambda_{n}}^{p_{1}\!\dots\! p_{n}}b_{\varphi}(\{\lambda_{k}\})|_{\lambda_{i}=0} =\left\langle b(\rho_{1}\!\dots\!\rho_{n})\rho_{1}^{p_{1}}\!\dots\!\rho_{n}^{p_{n}}\right\rangle_{c}$.
The first term in equation~(\ref{eq:Cpp}) is the error on the mean which is the typical error if draws are independent. The correlations between the draws  -- i.e the cells -- lead to an additional source of errors encoded in the second term which corresponds to the bias function. 
Note that as expected, in the very large separation limit where $\xi(r_\mathrm{IJ}) \rightarrow \delta_{\mathrm{IJ}} $,
we get
$C_{\mathbf p \mathbf q} \rightarrow  C^0_{\mathbf p \mathbf q}/\mathrm{N_{t}}$ {\red where $C^0_{\mathbf p \mathbf q}= \left\langle \rho_{1}^{p_{1}+q_{1}}\dots\rho_{n}^{p_{n}+q_{n}}\right\rangle_{c}$ and $\delta_{\mathrm{IJ}}$ is the Kronecker delta function here}. 

\subsection{Errors on the PDF}
\label{sec:errPDF}
Let us finally  quantify the cosmic variance on the estimate of the one-cell PDF when measuring densities in a finite number $\rm N_{t}$ of spheres.
In this case, it is necessary to take into account the discreteness of the counts and the size of the bins of density. 
%{\red All the details are given in Appendix~\ref{discretecounts}. We propose here to sum up our main findings in the Poisson limit.}

{\red One can show ({see Appendix~\ref{discretecounts} for details}) that in the Poisson limit,} the {\red number} $\rm N$ of spheres with density in the interval $ \Delta=[\rho-\Delta\rho/2,\rho+\Delta\rho/2]$ is unbiased ($\left\langle \rm N\right\rangle=\bar N$) and has variance
\begin{equation}
\label{eq:var1cell}
\left\langle \rm N^{2}\right\rangle-\left\langle \rm N\right\rangle^{2}={\rm \bar N}+b^{2}\xi\rm\bar N^{2}\,,
\end{equation}
where $\xi$ is the mean correlation between the spheres, $\xi=\sum_{\rm I\neq J}\xi(r_{\rm IJ})/[\rm N_{t}(N_{t}-1)]$,
and  ${\rm \bar N}=p \rm N_{t}$  with  $
p=\int_{\Delta}\dd\rho\mP(\rho)
$ is the expected number of spheres with density in the interval considered.
For a large enough number of spheres, sampling errors can therefore be neglected and the cosmic variance is directly proportional to $b^{2}$ where here $b$ is defined as 
the mean density bias in the bin
\begin{equation}
b=\int_{\Delta}\mP(\rho)b(\rho)\dd\rho/\int_{\Delta}\mP(\rho)\dd\rho\,, \label{eq:defb}
\end{equation}
where the density bias, $b(\rho)$, entering equation~(\ref{eq:fullPDFlargeseparation}) will later be shown to obey equation~(\ref{eq:defbiasPDF}).

Similarly, the correlations between the counts in different bins of density can be investigated.
The {\red numbers} $\rm N_{1}$ of spheres with density in $ \Delta_{1}=[\rho_{1}-\Delta\rho/2,\rho_{1}+\Delta\rho/2]$ and $\rm N_{2}$ of spheres with density in $\Delta_{2}=[\rho_{2}-\Delta\rho/2,\rho_{2}+\Delta\rho/2]$  are unbiased and have a covariance
\begin{equation}
\label{eq:var2cell}
\left\langle \rm N_{1}N_{2} \right\rangle={\rm \bar N_{1}\bar N_{2}}(1+\xi b_{1}b_{2})\,,
\end{equation} 
with $b_{1}$ and $b_{2}$ being  defined as in equation~(\ref{eq:defb}) for the bins $\rho_{1}\pm\Delta\rho/2$ and $\hat\rho_{2}\pm\Delta\rho/2$.
The proof of this result is again derived in Appendix~\ref{discretecounts} and can be easily generalized to any number of concentric cells.
%Note that equations~(\ref{eq:var1cell}) and (\ref{eq:var2cell}) are only valid in the Poisson limit (see Appendix~\ref{discretecounts} for the exact expression).

The consequence for the error budget of the one-cell PDF is as follows. Let us define $\hat\mP(\rho_{i})={\rm N_{i}/N_{t}}/\Delta\rho$, the estimate of the PDF measured from a set of $\rm N_{t}$ spheres when the range of densities is divided in bins centred on $\rho_{i}$ with width $\Delta \rho$. First, this estimator, $\hat\mP$, is unbiased.
Equations~(\ref{eq:var1cell}) and (\ref{eq:var2cell}) also yield the expected error on the estimate of the PDF
\begin{equation}
\left\langle \hat\mP(\rho_{i})^{2}\right\rangle-\left\langle \rm \hat\mP(\rho_{i})\right\rangle^{2}=\frac{\bar \mP(\rho_{i})}{\Delta\rho \rm N_{t}}+b_{i}^{2}\xi \left(\bar \mP(\rho_{i})\right)^{2}\,,
\label{autocorr}
\end{equation}
where the mean PDF in the bin is
$\bar \mP(\rho_{i})=\int_{\Delta_{i}}\mP(\rho)\dd\rho/\Delta\rho$ and $b_{i}^{2}\left(\bar \mP(\rho_{i})\right)^{2}$ is the mean value squared of the bias in the density bin, $\left(\int_{\Delta_{i}}\mP(\rho)b(\rho)\dd\rho/\Delta\rho\right)^{2}$.
Furthermore, the typical correlation between two distinct bins, $i\neq j$,  is given by
\begin{equation}
\label{eq:Cpp-onecell}
\left\langle \hat\mP(\rho_{i})\hat\mP(\rho_{j}) \right\rangle=\bar \mP(\rho_{i})\bar \mP(\rho_{j})(1+\xi b_{i}b_{j})\,.
\end{equation} 
In particular, it is straightforward to see that equation~(\ref{eq:Cpp-onecell}) is fully consistent with  equation~(\ref{eq:Cpp}) in the one-cell case.
Indeed, from equations~(\ref{eq:Cpp-onecell}) and ~(\ref{autocorr}), one can compute for instance the correlation between the estimated moment of order $p$ of the density and the moment of order $q$\begin{align}
M_{pq}&\!=\!\left\langle \overline{\rho^{p}}\, \overline{\rho^{q}}\right\rangle,\nonumber\\
&\!=\!\sum_{i,j}(\Delta\rho)^{2} \left\langle\hat\mP(\rho_{i})\hat\mP(\rho_{j})\right\rangle \rho_{i}^{p}\rho_{j}^{q},\nonumber\\
&\!=\!\sum_{i,j}(\Delta\rho)^{2} \bar\mP(\rho_{i})\bar\mP(\rho_{j})(1\!+\!\xi b_{i}b_{j}) \rho_{i}^{p}\rho_{j}^{q}\!+\!\!\sum_{i}\!\!\frac{\Delta\rho}{\rm N_{t}} \rho_{i}^{p+q}\bar \mP(\hrho_{i}),\nonumber\\
&\!=\!\frac 1 {\rm N_{t}}\left\langle\rho^{p+q}\right\rangle+\left\langle\rho^{p}\right\rangle\left\langle\rho^{q}\right\rangle
+\xi\left\langle b(\rho)\rho^{p}\right\rangle\left\langle b(\rho)\rho^{q}\right\rangle,
\nonumber
\end{align}
which in terms of cumulants can be rewritten as
\begin{equation}
C_{pq}=\left\langle \overline{\rho^{p}}\, \overline{\rho^{q}}\right\rangle_{c}=\frac 1 {\rm N_{t}}\left\langle\rho^{p+q}\right\rangle_{c}+\xi\left\langle b(\rho)\rho^{p}\right\rangle_{c}\left\langle b(\rho)\rho^{q}\right\rangle_{c}\,,
\end{equation}
so that equation~(\ref{eq:Cpp}) is recovered in the one-cell case.

  The rest of the paper is devoted to demonstrating and validating equation~(\ref{eq:fullPDFlargeseparation}).
  
%%%%%%%%%%
\section{Tree order generating function}
\label{sec:cumulant}

 Let us first  compute the generating function, $\varphi_b$ of the cumulants containing any power of the densities in one location and one power of the density at an arbitrary distance. 
 As we shall see, such cumulants enter the derivation of equation~(\ref{eq:fullPDFlargeseparation}).
 For that purpose, we first consider $n$ concentric cells in one location of space.

%%%%%%%%%%
\subsection{Definitions and relation to spherical collapse}

 \cite{Bernardeau14} (hereafter BPC)  computed $\mP(\{\rho_{k}\})$, 
 the joint one-point PDF of the density within concentric spheres, in a highly symmetric configuration (spherical symmetry) where non-linear solutions to the gravitational dynamical equations are known explicitly. 
The corresponding symmetry implies that the most likely dynamics (amongst all possible mappings between the initial and final density field) 
is that corresponding to  spherical  collapse.
In the limit of small variance, BPC showed using a saddle approximation that the  Laplace transform of  $\mP(\{\rho_{k}\})$ corresponds to the  cumulant generating function of densities in 
concentric cells  $\varphi(\{\lambda_{k}\})$, and can be predicted analytically.  This function is  indeed closely related to the non-linear evolution of a spherically symmetric perturbation in the linear growing mode regime and reads
\begin{equation}
\varphi(\{\lambda_{k}\})=\sum_{p_{i}=0}^{\infty}\ \langle \Pi_{i}\ {\rho_{i}}^{p_{i}}(R_{i})
\rangle_{c}\frac{\Pi_{i}\lambda_{i}^{p_{i}}}{\Pi_{i}p_{i}!}\,,
\label{phidef}
\end{equation}
where $\rho_{i}$ is the density (in units of the average density) within the radius $R_{i}$. For this 
construction, it is essential that the cells are all spherical and concentric.

Let us denote $\zeta(\tau)$ the non-linear transform of the density so that
\begin{equation}
\rho=\zeta(\tau)\,, 
\label{eq:rho2tau}
\end{equation}
where $\rho$ is the density within the radius $R$ and $\tau$ is the linear density contrast within the radius $R\rho^{1/3}$ (for mass conservation). 
An explicit possible fit for $\zeta(\tau)$ is given by
\begin{equation}
\zeta(\tau)=\frac{1}{(1-\tau/\nu)^{\nu}}\,, \label{eq:spherical-collapse}
\end{equation}
where $\nu$ can be adjusted to the actual values of the cosmological parameters ($\nu=21/13$ provides 
a good description of the spherical dynamics for an Einstein-de Sitter background for the range of $\tau$
values of interest).
The main result of  BPC was that the cumulant generating function at tree order could be computed 
explicitly from its Legendre transform\footnote
{
Recently, \cite{LDPinLSS} showed that the prediction for the cumulant generating function given by equation~(\ref{eq:phi2psi}) originates from a regime of large deviations \citep[see][for a review]{2011arXiv1106.4146T} at play in the gravitational evolution of cosmic structures.
}, $\Psi(\{\rho_{k}\})$, as
\begin{equation}
\varphi(\{\lambda_{k}\})=\sum_{i}\lambda_{i}\rho_{i}-\Psi(\{\rho_{k}\})\,, \label{eq:phi2psi}
\end{equation}
where the $\{\rho_{k}\}$ are functions of the $\{\lambda_{i}\}$ 
via  the stationary conditions
\begin{equation}
\lambda_{i}=\frac{\partial}{\partial\rho_{i}}\Psi(\{\rho_{k}\})\,, \quad {i=1\!\cdots\! n}\,.
\label{stationary1}
\end{equation}
Note that this condition can only be inverted as long as $\det \partial_{i}\partial_{j}\Psi\neq 0$ which defines a critical $n$-dimensional surface\footnote{\red 
The critical surface defined by $\det \partial_{i}\partial_{j}\Psi =0$ is at finite distance from the origin ensuring that the
 cumulant generating function has a non-zero radius of convergence.  As a consequence, and contrary to
what happens in the case of single or multivariate log-normal distributions \citep{1991MNRAS.248....1C,2011ApJ...738...86C,2012ApJ...750...28C,2015arXiv150804838C}, the density PDF here decays exponentially and Carleman's criterion ensures that it is uniquely 
defined from its moments \citep[see for instance][]{akhiezer1965classical}.}
Under the above mentioned assumptions the  rate function 
$\Psi(\{\rho_{k}\})$ entering equation~(\ref{eq:phi2psi}) is explicitly  written in term of the initial conditions as 
\begin{equation}
\Psi(\{ \rho_{k}\})=\frac{1}{2}\sum_{ij}\Xi_{ij}(\{ \rho_{k} \})\,\tau_{i}(\rho_{i})\tau_{j}(\rho_{j})\,,
\label{PsiDef}
\end{equation}
where $\Xi_{ij}(\{R_{k}\rho_{k}^{1/3} \})$ is the inverse matrix of $\Sigma_{ij}=\sigma^{2}(R_{i}\rho_{i}^{1/3},R_{j}\rho_{j}^{1/3})$, the initial cross correlation matrix of the densities 
computed at radii $R_{i}\rho_{i}^{1/3}$ and $R_{j}\rho_{j}^{1/3}$. The covariance matrix $\sigma_{ij}$ encodes all dependency with respect to the initial power spectrum.

Note that equation~(\ref{eq:phi2psi}) can be inverted as
\begin{equation}
\Psi(\{\rho_{k}\})=\sum_{i}\lambda_{i}\rho_{i}-\varphi(\{\lambda_{k}\})\,,
\end{equation}
with 
\begin{equation}
\rho_{i}=\frac{\partial}{\partial\lambda_{i}}\varphi(\{\lambda_{k}\})\,, \quad {i=1\!\cdots n}\,.
\label{stationary2}
\end{equation}
The stationary condition then gives the expression of $\rho_{i}$ as a function
of the variables $\{\lambda_{k}\}$. Such a solution can be equivalently expressed 
in terms of the corresponding values of $\tau_{i}$.

\begin{figure*}
\centering
\includegraphics[width=1.9\columnwidth]{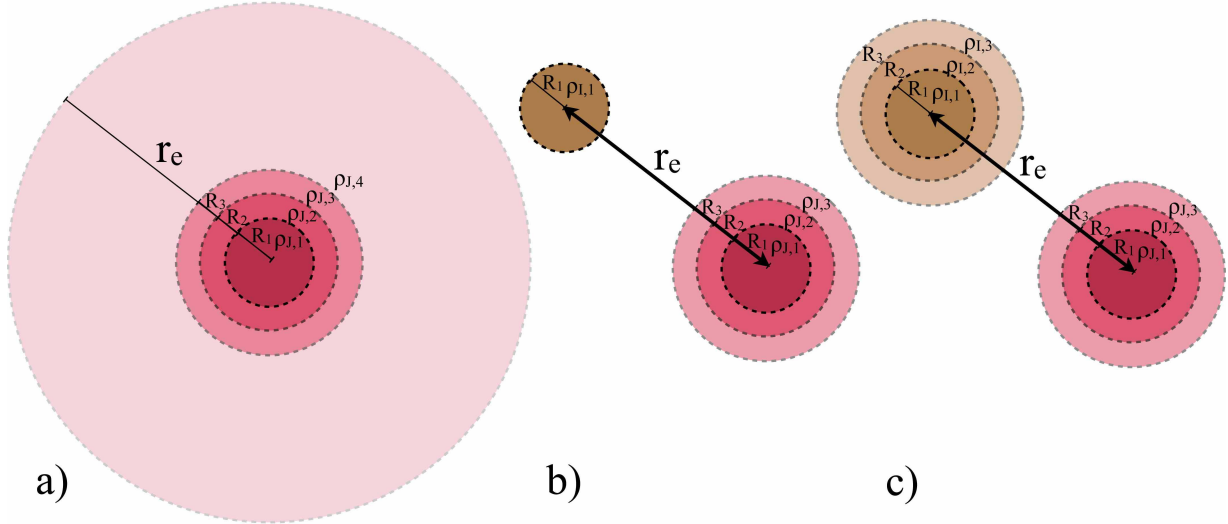}
   \caption{{\sl a)} The configuration of  $3+1$ spherical cells in one location (red cells shown in the left-hand panel) can be used to compute the joint cumulants involving any power of the density in 3 concentric cells in one location  (red cells displayed in the  {\sl b) panel}) and one power of the density in one cell (coloured in brown in the middle panel) at some arbitrary distance $r_{e}$ from the rest, as described in Section~\ref{sec:nplus1}. Those cumulants are the building blocks of the two-point PDF of concentric densities in the large-separation limit (see equation~(\ref{eq:cummixed})). The corresponding configuration  with $n=3$ concentric cells in one location (red) and $m=3$ concentric cells at a distance $r_{e}$ (brown) is displayed in the {\sl c) panel}.
   \label{fig:config-nplus1}}
\end{figure*}

\subsection{The n+1 cell formalism}
\label{sec:nplus1}

Let us now consider the formal derivation of $\varphi_{b}$, the {``one external leg''} generating function of joint cumulants for $n+1$ cells centred on the same point
 when the $n+1$th radius, $R_{n+1}=r_{e}$, 
is set apart (at this stage there is no assumption on the relative size of these radii). This configuration is illustrated in the left-hand panel of Fig.~\ref{fig:config-nplus1}
and is of particular interest since we will show in this section how it can be used to predict some configurations of the two-point statistics without any assumption on the separation. Later, we will also use it as a building block of the large-separation approximation of the two-point correlation function of concentric densities (see Section~\ref{sec:np}). This generating function simply reads
\begin{equation}
\varphi_{b}(\{\lambda_{k}\};<r_{e})=\sum_{p_{i}=0}^{\infty}\ \langle \rho(r_{e}) \Pi_{i=1}^{n}\ {\rho_{i}}^{p_{i}}(R_{i})
\rangle_{c}\frac{\Pi_{i=1}^n\lambda_{i}^{p_{i}}}{\Pi_{i=1}^n p_{i}!}\,,
\label{phibdef}
\end{equation}
where $\rho(r_{e})$ enters the cumulant $\langle \rho(r_{e}) \Pi_{i=1}^{n}\ {\rho_{i}}^{p_{i}}(R_{i})
\rangle_{c}$ only as a linear power. 
Equation~(\ref{phibdef}) is the generating function of the cumulants containing one power of the outer density and arbitrary powers of the $n$ inner densities.
It simply corresponds to the first derivative of the cumulant generating functions for $n+1$ cells taken at the origin
\begin{equation}
\varphi_{b}(\{\lambda_{k}\};<r_{e})=\frac{\partial }{\partial \lambda_{n+1}}\varphi(\lambda_{1},\dots,\lambda_{n+1})\Big\vert_{\lambda_{n+1}=0}\,.
\end{equation}
Taking advantage of the stationary condition (\ref{stationary2}) applied to $\lambda_{n+1}$, we also have
\begin{equation}
\varphi_{b}(\{\lambda_{k}\};<r_{e})=\rho_{n+1}(\lambda_{1},\dots,\lambda_{n},0)\,, \label{eq:varphib}
\end{equation}
where $\rho_{n+1}$ is in turn computed in terms of the $\lambda_{i}$ from the set of stationary conditions~(\ref{stationary1}).
Finally, equation~(\ref{eq:varphib}) can also be re-expressed via equation~(\ref{eq:rho2tau}) in terms of the corresponding linear density contrast  as
\begin{equation}
\varphi_{b}(\{\lambda_{k}\};<r_{e})=\zeta\left(\tau(r_{e})\right)\,, \label{eq:phiboftau}
\end{equation}
where $\tau(r_{e})\equiv \tau_{n+1}(\lambda_1,\cdots, \lambda_n,0)$
is to be computed as a function of $\{\lambda_{k}\}$ for the specific case where $\lambda_{n+1}$ is set to 0.
We can then take advantage of decimation (see Appendix~\ref{decimation}, equation~(\ref{eq:taure}))
to write  $\tau(r_{e})$ via  the  implicit equation
\begin{align}
\tau(r_{e})=&\sum_{i=1}^{n}\sigma^{2}(r_{e}\,\zeta(\tau(r_{e}))^{1/3},R_{i}\,\zeta(\tau_{i})^{1/3}) \times\nonumber\\
&\sum_{j=1}^{n}\Xi_{ij}(\{R_{k}\,\zeta(\tau_{k})^{1/3})\})\tau_{j}\,,
\label{tauRe}
\end{align}
where the tensor and vector quantities ($\Xi_{ij}$, $\tau_{i}$) are computed when only the first $n$ cells are
considered  (so that the set $\sum_{j=1,n}\Sigma_{ij}\Xi_{jk}=\delta_{ik}$ together with the stationary conditions form
a set of $n$ coupled equations only). Technically,  
equation~(\ref{tauRe}) can be solved given the values of $\{\tau_{k}\}_{k=1, \cdots\!, n}$ which in turn can be expressed in terms
of the variables $\{\lambda_{k}\}_{k=1, \cdots\!, n}$.

Now note that  equations~(\ref{eq:phiboftau}) and (\ref{tauRe}) can be used to get the cumulant generating functions for any quantities linearly related to the density. In particular,
 the density in an infinitesimal  shell
at a given distance $r_e$ reads $\rho(r_{e}<r<r_{e}+\dd r_{e})=\dd \rho(r_{e}) r_{e}^{3}/\dd r_{e}^{3}$ so that the corresponding cumulant generating function, $\varphi_{b}(\{\lambda_{k}\};r_{e})$,
 can be written as
\begin{equation}
\varphi_{b}(\{\lambda_{k}\};r_{e})=\frac{1}{r_{e}^{2}}\frac{\dd}{\dd r_{e}}\left(\frac{r^{3}_{e}}{3}\varphi_{b}(\{\lambda_{k}\};<r_{e})\right)\,. \label{eq:defphibre}
\end{equation}
Thanks to rotational invariance, the value of the cumulant within an infinitesimal  shell
at a distance $r_e$ is the same as if the density was computed at a distance $r_e$ in any direction. 
Therefore, equation~(\ref{eq:defphibre})
also describes the cumulant generating function of concentric densities in spheres of radii $R_{i}$ ($1\leq i\leq n$) and density at some given distance $r_{e}$.

Finally note that the domain for $\rho_{e}$ does not need to be a spherical cell and
equation~(\ref{eq:defphibre})
 can subsequently be integrated in any domain $\mS$ {of arbitrary shape}
\begin{equation}
\varphi_{b}(\{\lambda_{k}\};\mS)=\frac{1}{V_{\mS}}\int_{\mS}{\dd^{3}\vecr}\ \varphi_{b}(\{\lambda_{k}\};r)\,,
\end{equation}
with $V_{\mS}$  the volume of the domain. 
This configuration is illustrated in the middle panel of Fig.~\ref{fig:config-nplus1}.

\section{Bias in the large-separation limit}
\label{sec:bias}
In the large-separation limit, $r_e\gg R_{\mathrm {max}}=\max_{j}R_j$, the results of the previous section can be pursued further
 to investigate the effects of cosmic variance  on the measurement of the statistical properties of the field at scales $R_{j}$ within a much larger survey.

\subsection{Bias of the $n+1$ cumulant generating function}

In the large-separation limit where the internal radii, $R_{i}$, are all smaller than $r_{e}$ and for realistic power spectra on cosmological scales, the cross-correlations  $\sigma(r_{e}\,\zeta(\tau(r_{e}))^{1/3},R_{i}\,\zeta(\tau_{i})^{1/3})$ are much smaller than any internal moments $\sigma(R_{i}\,\zeta(\tau_{i})^{1/3},R_{j}\,\zeta(\tau_{j})^{1/3})$. This property\footnote{This hierarchy of moments can be investigated by varying the spectral index and radii in equation~(\ref{sigij}) and shown to be valid for the range of power spectra and radii of cosmological interest.} implies
that in equation~(\ref{tauRe}) the coefficients $\sigma^{2}(r_{e}\,\zeta(\tau(r_{e}))^{1/3},R_{i}\,\zeta(\tau_{i})^{1/3})\Xi_{ij}$ are smaller that unity and therefore $\tau(r_{e})$ is also small. The leading order expression of $\varphi_{b}$ can then be obtained using a Taylor expansion around $\tau(r_{e})\approx 0$ of equation~(\ref{eq:spherical-collapse})
into equations~(\ref{eq:phiboftau}) and (\ref{tauRe}) so that
\begin{equation}
\varphi_{b}(\{\lambda_{k}\};<r_{e})=1+\!\!
\sum_{i=1}^{n}\!\sigma^{2}\big(r_{e},R_{i}
\,\zeta(\tau_{i})^{1/3}
\big)\!\!
\sum_{j=1}^{n}\Xi_{ij}\tau_{j}. \hskip -0.3cm \label{eq:varphilarge}
\end{equation}
Pursuing this approximation further,  we also expect that $\sigma(r_{e},R_{i}\,\zeta(\tau_{i})^{1/3})$ are essentially all equal 
and given by
\begin{equation}
\sigma(r_{e},R_{i}\,\zeta(\tau_{i})^{1/3})\approx \sigma(r_{e},0)\equiv \sigma(<r_{e}).
\end{equation}
 Equation~(\ref{eq:varphilarge}) can then be simplified as
\begin{equation}
\varphi_{b}(\{\lambda_{k}\};<r_{e})=1+
\sigma^{2}(<r_{e}) \sum_{i=1}^{n} \sum_{j=1}^{n}\Xi_{ij}\tau_{j}\,.
\end{equation}
In particular, this implies, via equation~(\ref{eq:defphibre}), that the
 cumulant generating function for the density at some large distance $r_{e}$ from the $n$ cells obeys
\begin{equation}
\varphi_{b}(\{\lambda_{k}\};r_{e})=1+
\xi(r_{e}) \sum_{i=1}^{n} \sum_{j=1}^{n}\Xi_{ij}\tau_{j}\,,
\end{equation}
where $\xi(r_{e})$ is the dark matter correlation function at distance $r_{e}$
\begin{equation}
\xi(r_{e})\equiv\frac{1}{r_{e}^{2}}\frac{\dd}{\dd r_{e}}\left(\frac{r^{3}_{e}}{3}\sigma^{2}(<r_{e})\right)\,.
\end{equation}
Let us then {\sl define} the bias {\red cumulant generating function}
\begin{equation}
b_{\varphi}(\{\lambda_{k}\})\equiv\sum_{i=1}^{n} \sum_{j=1}^{n}\Xi_{ij}\tau_{j}\,, \label{eq:defbias}
\end{equation}
so that
 \begin{equation}
 \label{eq:bias-phib}
\varphi_{b}(\{\lambda_{k}\};r_{e})= 1+\xi(r_{e})b_{\varphi}(\{\lambda_{k}\})\,.
\end{equation}
Using equation~(\ref{eq:defbias}), the bias cumulant generating function can in principle be computed by means of equations~(\ref{eq:rho2tau}), (\ref{stationary1}) and the inverse of the  cross correlation matrix of the densities defined  near (\ref{PsiDef}) for any number of cells.

Within this approximation we see, recalling equation~(\ref{phibdef}), that
all cumulants of the form
$\langle \rho(r_{e})\,\rho_{1}^{p_{1}}\dots\rho_{n}^{p_{n}}
\rangle_{c}$, are  proportional to $\xi(r_{e})$,
where the $\rho_{i}(R_{i})$ and $\rho(r_{e})$ are located on cells centred at distance $r_{e}\gg R_i$ from one another.

\begin{figure}
\includegraphics[width=\columnwidth]{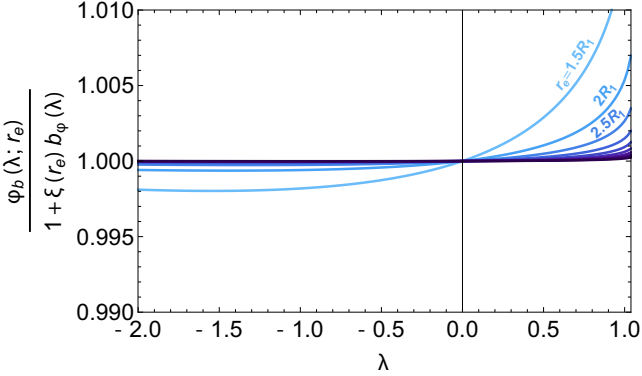}
   \caption{Ratio between $\varphi_{b}(\lambda;r_{e})$, the exact generating function of cumulants of the form
$\langle \rho_{1}^{p}\rho(r_{e})\rangle_{c}$ and its large-distance approximation $\varphi_{b}(\lambda;r_{e})= 1+\xi(r_{e})b_{\varphi}(\lambda)$ for a power-law density power spectrum with index $n_{s}=-1.6$ and variance $\sigma^{2}=0.3$. The separation $r_{e}$ spans the range between $r_{e}=3/2R_{1}$ (light blue) and $r_{e}=6R_{1}$ (dark blue) as labelled. Note that the critical point in this case is $\lambda_{c}\approx1.05$.
   \label{fig:phib}}
\end{figure}

For $n=1$ cell, Fig.~\ref{fig:phib} shows the convergence of the approximation defined by equation~(\ref{eq:bias-phib}) towards the exact value of $\varphi_{b}(\lambda;r_{e})$ computed from equations~(\ref{eq:spherical-collapse}), (\ref{eq:phiboftau}) and (\ref{tauRe}) as the distance $r_{e}$ increases. The convergence is fast,  reaching sub-percent precision even for relatively small separations $r_{e}\gtrsim2R_{1}$. As expected, the agreement is best for small values of $\lambda$, while deviations appear in the tails. Those conclusions are expected to hold similarly for a higher number $n$ of cells.

\subsection{The n+m formalism at  large separation }
\label{sec:np}
Let us finally focus on the $n+m$ formalism where $n$ concentric cells at a distance $r_{e}$ from $m$ other concentric cells are considered. 
The rest of the paper is devoted to the study of this configuration,  illustrated in the right-hand panel of Fig.~\ref{fig:config-nplus1}, in order to investigate the large-separation two-point statistics of densities in concentric cells.

In  this large-separation limit,
 we can  write down (see appendix \ref{pkbgarg}) the joint cumulants for two sets, separated by $r_e$ much larger than $R_{i}$,
 at leading order 
in terms of the two-point cumulants containing one power of the density in one location and any powers of the density in the second location (and generated by $\varphi_{b}$, see Section~\ref{sec:nplus1})
\begin{align}
&\langle 
\rho_{1}^{p_{1}}\dots\rho_{n}^{p_{n}}
{\rho'}_{1}^{q_{1}}\dots{\rho'}_{m}^{q_{m}}
\rangle_{c}=
\nonumber\\&\hspace{1cm}
\frac{1}{\xi(r_{e})}
\langle 
\rho_{1}^{p_{1}}\dots\rho_{n}^{p_{n}}
{\rho'}_{i}
\rangle_{c}
\langle 
\rho_{i} {\rho'}_{1}^{q_{1}}\dots{\rho'}_{m}^{q_{m}}
\rangle_{c}\,, \label{eq:cummixed}
\end{align}
where the $\{\rho_k=\rho_{1,k}\}_{k}$ correspond to a set of radii $\{R_k\}_{k}$ of cells  centred at, say, the origin, the $\{\rho'_k=\rho_{2,k}\}_{k}$  correspond
to a set of radii $\{R'_k\}_{k}$ of cells centred on a point at distance $r_{e}$ from the origin and $p_{i}$ and  $q_{i}$ are non-zero. 
Equation~(\ref{eq:cummixed}) can be easily understood from a diagrammatic point of view. Indeed, in the large-separation regime, the dominant contribution will come from the configuration in which there is only one leg linking two connected diagrams belonging to each location (see also \cite{1996A&A...312...11B} for details).

At the level of the generating functions, equations~(\ref{eq:bias-phib}) and (\ref{eq:cummixed}) imply that
\begin{align}
&\varphi(\{\lambda_{k}\},\{\lambda'_{k}\};r_{e})= 
\nonumber\\
&\hspace{.4cm}\varphi(\{\lambda_{k}\}){+} \varphi(\{\lambda'_{k}\}) +
\xi(r_{e})\, b_{\varphi}(\{\lambda_{k}\}) \, b_{\varphi}(\{\lambda'_{k}\}) \,, \label{eq:phirhokrhokp}
\end{align}
where
$\varphi(\{\lambda_{k}\},\{\lambda'_{k}\};r_{e})$ is the generating functions of the joint cumulants
$\langle 
\rho_{1}^{p_{1}}\dots\rho_{n}^{p_{n}}
{\rho'}_{1}^{q_{1}}\dots{\rho'}_{m}^{q_{m}}
\rangle_{c}$.
Equation~(\ref{eq:phirhokrhokp}) is the cornerstone of this paper.

\subsection{Consequences for the joint PDFs}

The structure of equation~(\ref{eq:phirhokrhokp}) for the cumulant generating function has direct consequences at the 
level of the  corresponding joint PDF.
Let us  consider again two  sets of concentric cells separated by a 
distance $r_e$, 
and define the corresponding densities $\{\hrho_{k}\}\equiv\{\hrho_{1,k}\}$ and $\{\hrho'_{k}\}\equiv\{\hrho_{2,k}\}$.
For large separations, the joint PDF $\mP(\{\hrho_{k}\},\{\hrho'_{k}\};r_{e})$ takes, at leading order, the following form 
\begin{multline}
\mP(\{\hrho_{k}\},\{{\hrho'}_{k}\};r_{e})=\\
\mP(\{\hrho_{k}\})\mP(\{{\hrho'}_{k}\})
\left[1+\xi(r_{e})
b(\{\hrho_{k}\})b(\{{\hrho'}_{k}\})\right]\,, \label{eq:jointPDF}
\end{multline}
given that
\begin{equation}
 \label{eq:PDF}
\mP(\{\hrho_{k}\})=
\int \frac{\dd \lambda_{1}}{2\pi \ii} \dots \frac{\dd \lambda_{n}}{2\pi \ii} \exp\left({ -}\lambda_{i}\hrho_{i}{ +}\varphi(\{\lambda_{k}\})\right)\,,
\end{equation}
and
\begin{multline}
b(\{\hrho_{k}\})\, \mP(\{\hrho_{k}\})=\\
\int \frac{\dd \lambda_{1}}{2\pi \ii} \dots \frac{\dd \lambda_{n}}{2\pi \ii} 
b_{\varphi}(\{\lambda_{k}\})
\exp\left({ -}\lambda_{i}\hrho_{i}{ +}\varphi(\{\lambda_{k}\})\right),  \label{eq:defbiasPDF}
\end{multline}
where the bias {\red cumulant generating function} $b_{\varphi}(\{\lambda_{k}\})$ is given by equation~(\ref{eq:defbias}) and we have introduced the corresponding effective bias function, $b(\{\hrho_{k}\})$ (that will also simply be called ``bias'' or ``bias function'' in what follows). {\red  Note that measured densities will henceforth be denoted with a hat in order to avoid confusion with variables intervening in the computation of the cumulant generating functions (see for instance equation~(\ref{eq:phi2psi})).}

Equation~(\ref{eq:defbiasPDF}) is one of the main results of this paper.
It defines the bias functions we introduced in equation~(\ref{eq:fullPDFlargeseparation}),
which is dual to  equation~(\ref{eq:defbias}) in the duality defined by PDF versus generating functions. Correspondingly, equation~(\ref{eq:jointPDF}) is dual to equation~(\ref{eq:phirhokrhokp}).
It is easy to show that the bias functions, $b(\{\hrho_{k}\})$, have an analytical asymptote at low density which can be derived using a steepest descent method in equations~(\ref{eq:PDF}) and (\ref{eq:defbiasPDF})
\begin{equation}
\label{eq:blim}
b(\{\hrho_{k}\})\approx b_{\varphi}(\{\lambda_{k}=\partial_{k}\Psi(\{\hrho_{i}\})\})\,,
\end{equation}
where $\partial_{k}\Psi=\partial \Psi/\partial \rho_{k}$. In practice, we will make use below of the following low-density approximation
\begin{equation}
\label{eq:blim2}
b(\{\hrho_{k}\})\approx \sum_{i=1}^{n} \sum_{j=1}^{n}\Xi_{ij}(\{R_{k}\hrho_{k}^{1/3}\})
\zeta^{-1}(\hrho_{j})\,,
\end{equation}
which allows to go beyond the critical point where $\lambda_{i}=\partial_{i}\Psi(\{\hrho_{k}\})$ is ill-defined.

\subsection{Properties of the density bias}
\label{sec:properties}
From equation~(\ref{eq:defbias}), one can show that $b_{\varphi}=0$ and $b_{\varphi}'(0)=1$. Using equation~(\ref{eq:defbiasPDF}), it follows that the bias functions obey the following two relations
\begin{align}
& \int_{0}^{\infty}\dd \hrho_{k} \mP(\{\hrho_{k}\})\,b(\{\hrho_{k}\})=0\,,\\
&\int_{0}^{\infty}\dd \hrho_{k} \mP(\{\hrho_{k}\})\,b(\{\hrho_{k}\}) \hrho_{k}=1\,.
\end{align}
These properties ensure the normalisation of the PDF and the definition of $\xi$, the dark matter correlation function.

In particular, we will make use of these identities to measure the density bias function either using the auto-correlation of cells of a given density
\begin{multline}
1+b^{2}(\hrho)\xi(r_{e})
=\\
\frac{\displaystyle\int_{0}^{\infty}\dd \hrho_{k}\dd \hrho'_{k} \mP(\{\hrho_{k}\},\{{\hrho'}_{k}\};r_{e})\delta_{\text D}(\hrho_{k}-\hrho)\delta_{\text D}(\hrho'_{k}-\hrho)}{\displaystyle\int_{0}^{\infty}\dd \hrho_{k}\dd \hrho'_{k} \mP(\{\hrho_{k}\})\mP(\{\hrho'_{k}\})\delta_{\text D}(\hrho_{k}-\hrho)\delta_{\text D}(\hrho'_{k}-\hrho)}
\label{eq:brho-auto}\,,
\end{multline}
or their cross-correlations
\begin{multline}
\label{eq:brho-cross}
1+b(\hrho)\xi(r_{e})=
\\
\frac{\displaystyle\int_{0}^{\infty}\dd \hrho_{k}\dd \hrho'_{k} \mP(\{\hrho_{k}\},\{{\hrho'}_{k}\};r_{e})\delta_{\text D}(\hrho_{k}-\hrho)\hrho'_{k}}
{\displaystyle\int_{0}^{\infty}\dd \hrho_{k} \mP(\{\hrho_{k}\})\delta_{\text D}(\hrho_{k}-\hrho)}\,.
\end{multline}
In practice, measurements will be done in bins of a given width, meaning that Dirac delta functions will be replaced by stepwise functions (see Section~\ref{sec:measure-density-bias} below).

\subsection{Slope bias}
In the two-cell case,  instead of $(\hrho_{1},\hrho_{2})$, we will use the variables $(\hrho,\hat s)$ describing the inner density $\hrho=\hrho_{1}$ and slope $\hat s=(\hrho_{2}-\hrho_{1})R_{1}/\Delta R$ with $\Delta R=R_{2}-R_{1}$. The slope bias, $b(\hat s)$, will be investigated using the relation
\begin{equation}
\label{eq:srho-cross}
1+b(\hs)\xi(r_{e})
=
\frac{\displaystyle\int_{0}^{\infty}\!\!\!\!\!\dd \hrho\int_{-\infty}^{\infty}\!\!\!\!\!\dd \hs' \,\mP(\{\hrho\},\{\hs'\};r_{e})\delta_{\text D}(\hs' -\hs)\hrho}
{\displaystyle\int_{-\infty}^{\infty}\!\!\!\!\!\dd \hs'\, \mP(\hs')\delta_{\text D}(\hs'-\hs)}\,.
\end{equation}
In equation~(\ref{eq:srho-cross}), $\mP(\{\hrho\},\{\hs'\};r_{e})$ is a marginal of the two-cell PDF given by $\mP(\{\hrho_{1},\hs_{1}\},\{\hrho_{2},\hs_{2}\};r_{e})\delta_{\rm D}(\hrho_{1}-\hrho)\delta_{\rm D}(\hs_{2}-\hs')$ integrated over $ \hrho_{1}$, $ \hs_{1}$, $\hrho_{2}$ and $\hs_{2}$.
The slope bias, $b(\hs)$, is defined as the Inverse Laplace transform 
of the slope bias cumulant generating function 
\begin{equation}
b_{s}(\mu)\equiv b_{\varphi}\left(-\frac{R_{1}}{\Delta R}\mu,\frac{R_{1}}{\Delta R}\mu\right)\,,
 \label{eq:defbias-slope}
\end{equation}
via 
\begin{equation}
b(\hat s)\, \mP(\hat s)=
\int \frac{\dd \mu}{2\pi \ii} 
b_{s}(\mu)
\exp\left({ -}\mu\hat s{ +}\varphi_{s}(\mu)\right)\,,  \label{eq:defbiasPDF-slope}
\end{equation}
with the slope PDF defined as
\begin{equation}
 \mP(\hat s)=
\int \frac{\dd \mu}{2\pi \ii} 
\exp\left({ -}\mu\hat s{ +}\varphi_{s}(\mu)\right)\,.  \label{eq:defslopePDF}
\end{equation}
Equations~(\ref{eq:defbiasPDF-slope}) and (\ref{eq:defbias-slope}) follow from the rewriting
\begin{align}
\lambda_{1}\rho_{1}+\lambda_{2}\rho_{2}&=(\lambda_{1}+\lambda_{2})\rho_{1}+\frac{\Delta R}{R_{1}}\lambda_{2}s\,, \\
&=\lambda \rho +\mu s\,,
\end{align}
which makes explicit the relation between $(\lambda_{1},\lambda_{2})$ and $(\lambda,\mu)$ and consequently the relation between the slope cumulant generating function, $\varphi_{s}$, and $\varphi$
\begin{equation}
\label{eq:varphis}
\varphi_{s}(\mu)=\varphi\left(\lambda_{1}=-\frac{R_{1}}{\Delta R}\mu,\lambda_{2}=\frac{R_{1}}{\Delta R}\mu\right)\,.
\end{equation}
Imposing $\lambda=0$ therefore allows us to  marginalise over $\rho$ and get the slope bias.

It has to be noted that the slope bias obeys similar contraints as the density bias namely
\begin{align}
& \int_{-\infty}^{\infty}\dd \hs\, \mP(\hs)\,b(\hs)=0\,,\\
&\int_{-\infty}^{\infty}\dd \hs\, \mP(\hs)\,b(\hs) \hs=0\,.
\end{align}

\section{Implementation for realistic $P_k$}
%%%%%%%%%%
\label{sec:prediction}

\begin{figure*}
\includegraphics[width=\columnwidth]{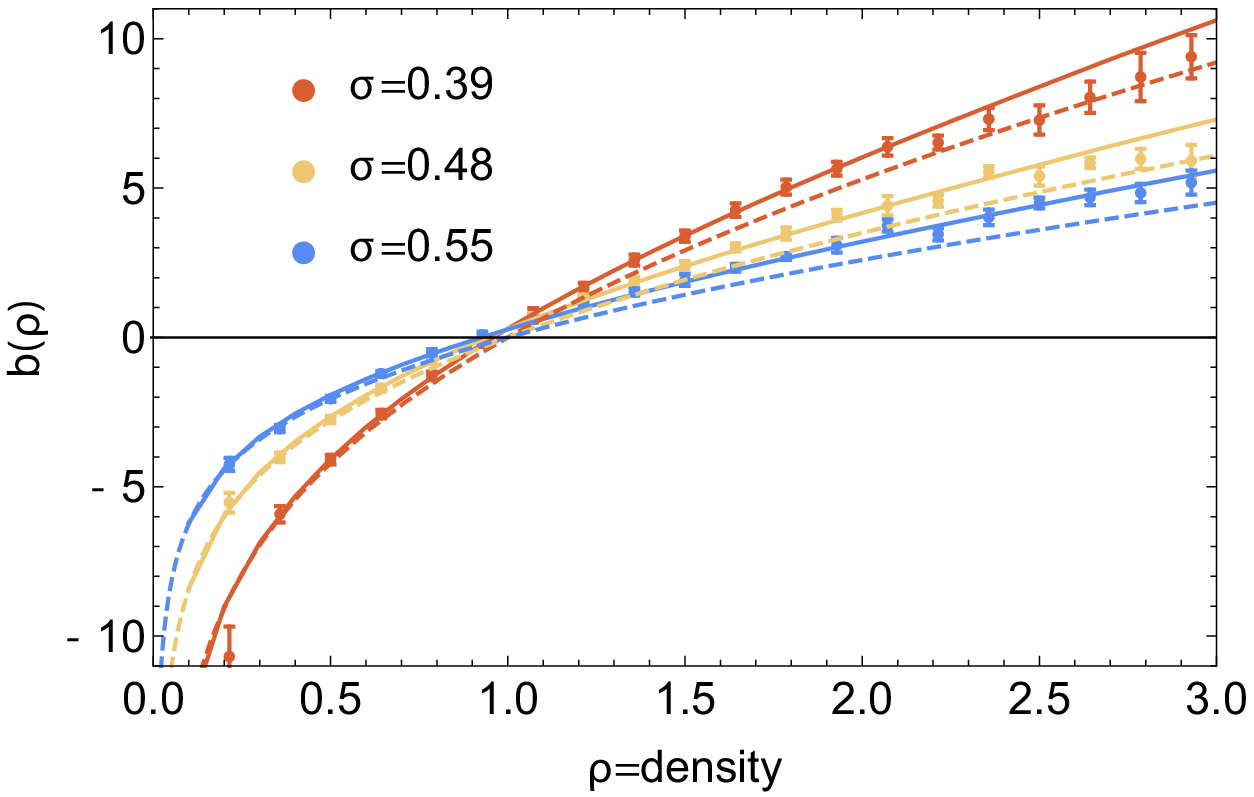}
\includegraphics[width=1.05\columnwidth]{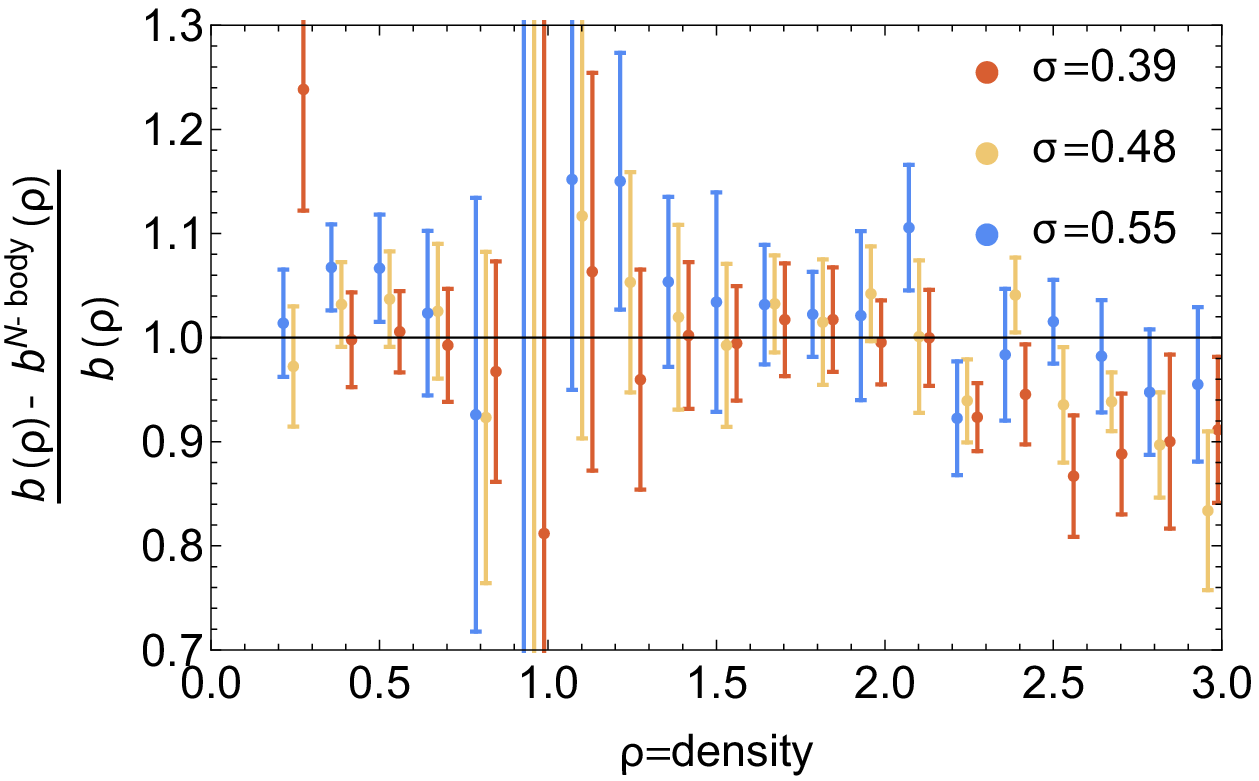}
   \caption{Left-hand panel: The density bias function $b(\hrho)$ for different values of the variance ($\sigma$=0.39, 0.48 and 0.55 for respectively the red, yellow and blue lines). The prediction computed numerically via equation~(\ref{eq:defbiasPDF}) is displayed with solid lines while the dashed lines correspond to the analytical low-density approximations given by equation~(\ref{eq:blim2}). Measurements in the simulation for spheres of radius $R=10$Mpc$/h$ separated by $r_{e}=20$Mpc$/h$ are shown with error bars and are successfully compared to the numerical prediction (solid lines) for the full range of variances and densities. The low-density approximations break down for $\rho\gtrsim 1$. Right-hand panel: Residuals between the measured density bias (error bars) and its numerical prediction from equation~(\ref{eq:defbiasPDF}) (displayed with solid lines on the left-hand panel). The yellow and red error bars have been shifted along the x-axis respectively by 0.03 and 0.06 for aesthetic purposes.
   \label{bias-rho-th}
}
\end{figure*}

In order to explicitly compute the bias functions for a realistic ($\Lambda$CDM) power spectrum, $P^{\lin }(k)$, and compare those predictions to simulations, we follow BPC and approximate the variance 
\begin{equation}
\sigma^{2}(R,R)=\int\frac{\dd^{3}\vk}{(2\pi)^{3}}P^{\lin }(k)W^{2}_{\rm 3D}(k R)\,,
\end{equation}
using an analytical
formula 
describing the scale-dependencies 
\begin{equation}
\sigma^{2}(R,R)=\frac{2\sigma^2(R_{p},R_{p})}{(R/R_p)^{n_1+3}+(R/R_p)^{n_2+3}}\,, \label{eq:sigmaeff}
\end{equation}
where $n_{1}$ and $n_{2}$ are chosen to reproduce the linear theory index $n(R)=-3-{\dd\log(\sigma^{2}(R,R))}/{\dd\log R}$  and running, $\alpha(R)={\dd\log(n(R))}/{\dd\log R}$ at the pivot scale $R_{p}$. Note that in fine the amplitude, $\sigma_{0}=\sigma(R_{p},R_{p})$, can be simply measured and not necessarily predicted by linear theory as it can be easily scaled out using the relation
\begin{equation}\label{scaling}
\varphi_{\sigma_{0}}(\{\lambda_{k}\})=\frac{1}{\sigma_{0}^{2}}\varphi_{1}(\{\lambda_{k}/\sigma_{0}^{2}\}),
\end{equation}
while the variable $\rho_{k}$ are independent of $\sigma_{0}$.
In practice, we take here $(n_{1}+n_{2})/2=-1.58$ and $n_{1}-n_{2}=1.23$ at $R_{p}=10 $Mpc$/h$ and $\sigma^{2}(R_{p},R_{p})$ is measured at different redshifts in the simulation.

Using the parametrization (\ref{eq:sigmaeff}) of $\sigma(R,R)$, we can then analytically predict the one-cell rate function $\Psi$, the cumulant generating function $\varphi$ and the bias function $b_{\varphi}$ and finally numerically compute the PDF $\cal P(\hrho)$ and bias  function $b(\hrho)$. In this paper, the numerical integrations are done, in the one-cell case, using a path parallel to the imaginary axis $\lambda=\lambda_{0}+\Delta \lambda\, \ii$ where $\lambda_{0}={\Psi'}(\textrm{min}(\hrho,\rho_{c}))$, $i$ goes from 1 to the number of points $n_{\rm steps}$ which is set to 1000 and the step, $\Delta\lambda$, depends on the value of the density $\hat \rho$ (typically larger for smaller densities).

The resulting prediction for the density bias is shown in the left-hand panel of Fig.~\ref{bias-rho-th} (solid lines). 
As expected from previous studies, this figure shows that the bias is zero only in regions where the density is around the mean i.e $\rho\approx 1$. 
This can easily be understood as ${\cal P}(\{\hrho\},\{\hrho'=1\};r_{e})/{\cal P}(1)={\cal P}(\hrho)$ if $r_{e}$ is large enough. Overdense regions are then positively biased while underdense regions are less clustered (negatively biased).
This is also consistent with \cite{Kaiser84} as the left-hand panel of Fig.~\ref{bias-rho-th} shows that the larger the density, the stronger the bias. Note that unlike the Gaussian peak bias which scales like the contrast for large densities, here the scaling of the non-linear bias departs from linearity with typically $b\propto\rho^{0.8}$ for large densities but is recovered in the very large density limit $\hrho\gtrsim 10$. Indeed, it is found (see Appendix~\ref{sec:largedensity} for more details) that the asymptotic behaviour of the bias function is given by
\begin{equation}
b(\hrho)\xrightarrow[\hrho\rightarrow\infty]{}\frac {(\hrho-\rho_{c})
(\lambda_{c}\rho_{c}(\rho_{c}^{1/\nu}-1-1/\nu)+\frac{1}{\nu}\varphi_{c})
}{
\nu(\rho_{c}^{1/\nu}-1)^{2}\rho_{c}^{1-1/\nu}}
\end{equation} 
when the integral is dominated by the singular point in equations~(\ref{eq:PDF}) and (\ref{eq:defbiasPDF}).
It has to be noted that this asymptotic behaviour depends on the existence of the singular point, $\rho_{c}$, which is due to the application of a large-deviation principle to the density field.
However, as shown in \cite{uhlemann16}, this singular point can be removed by applying a large-deviation principle to a non-linear transformation of the density field, for instance its logarithm. In that case, we expect the asymptotic behaviour to be slightly modified as shown in that paper.

 Note also that
the bias
of large-density regions is reduced for larger variances. This trend persists even if 
the density bias is plotted against the contrast $\nu_{c}=(\hrho -1)/\sigma$ instead of $\hrho$. 
The left-hand panel of Fig.~\ref{bias-rho-th} also displays the low-density approximation given by equation~(\ref{eq:blim2}). This approximation is only valid in a limited range of densities typically for $\hrho\lesssim1$.

In the two-cell case, one need to compute the covariance matrix between initial densities in spheres of radii $R_{1}$ and $R_{2}$
\begin{equation}
\sigma^{2}(R_{i},R_{j})=\int\frac{\dd^{3}\vk}{(2\pi)^{3}}P^{\lin }(k)W_{\rm 3D}(k R_{i})W_{\rm 3D}(k R_{j})\,,
\end{equation}
where $W_{\rm 3D}$ is the top-hat filter function
\begin{equation}
W_{\rm 3D}(k)=\frac{3}{k^{2}}(\sin(k)/k-\cos(k))\,.
\end{equation}
Again, for the sake of simplicity, we choose to parametrize this covariance matrix by
\begin{align}
\sigma^{2}(R_{i},R_{i})&=\sigma^2(R_{p})\left(\frac{R_{i}}{R_{p}}\right)^{-n_{s}(R_{p})-3}\,,\\
\sigma^{2}(R_{i},R_{j> i})
&=\sigma^2(R_{p})R_{p}^{n_{s}(R_{p})+3}{\cal G}(R_{i},R_{j},n_{s}(R_{p}))\label{sigij}\,,
\end{align}
where
\begin{align}
{\cal G}(x,y,n_{s})&=R_{p}^{-n_{s}-3}\frac{\int{\dd^{3}\vk\,}k^{n_{s}}W_{\rm 3D}(k x)W_{\rm 3D}(k y)}{\int{\dd^{3}\vk\,}k^{n}W_{\rm 3D}(k R_{p})W_{\rm 3D}(k R_{p})}\nonumber
\\
&=
  \! \frac{ (x\!+\!y)^{\alpha} \!\! \left(\!x^2\!+\!y^2\!-\!\alpha x y\right)\!-\!(y\!-\!x)^{\alpha} 
 \!  \!\left(\!x^2\!+\!y^2\!+\!\alpha x y\right)}
   {2^{\alpha}(n_{s}+1) x^3 y^3  },\nonumber
\end{align}
with $\alpha=1-n_{s}$.
We are now in a position to compute the two-cell rate function, cumulant generating function, PDF and finally the two-cell bias function.
The slope bias $b(\hs)$ predicted by this formalism is shown in Fig.~\ref{sbias} (right-hand panel).
Two configurations are shown to be uncorrelated ($b(\hs)=0$) and roughly correspond to slopes $\hs\approx\pm0.5$ with some $\sigma$-dependence.
This situation is of particular interest as, according to equation~(\ref{autocorr}), cosmic variance is drastically reduced in this case where only subdominant contributions will appear such as Poisson noise and small-scale effects. Besides this noteworthy case, it is found that small slopes ($|\hs|\lesssim 0.5$) are negatively biased while regions with larger (positive or negative) slope are more clustered.
As expected, the bias is stronger for large slopes which typically correspond to  sharp peaks (or voids) and will scale like $b(\hs)\propto \hs$ in the very large slope limit as explained in appendix~\ref{sec:approx-slope}. 
This asymptotic large $|s|$ behaviour can be compared  once again to the linear Kaiser bias, given by equation~(\ref{eq:xikaiser}), but in 
   the contrast of the peak rather than the slope.
Note also that
the asymmetry of the bias function is weakened with variance.

The predictions for the density and slope bias functions are compared against simulations in the next section.

\section{Validation against simulations}
%%%%%%%%%%
\label{sec:validation}

 In the following  we present the measurement of this 
bias function for the density $\hat\rho$ and for the density slope $\hs$.

The  dark matter simulation  \citep[carried out with {\tt Gadget2},][]{gadget2} is characterized by the following $\Lambda$CDM cosmology: $\Omega_{\rm m}=0.265 $, $\Omega_{\Lambda}=0.735$, $n=0.958$, $H_0=70 $ km$\cdot s^{-1} \cdot $Mpc$^{-1}$ and $\sigma _8=0.8$, 
$\Omega_{b}=0.045$
within one standard deviation of WMAP7 results \citep{wmap7}. 
The box size is 500 Mpc$/h$ sampled with  $1024^3$ particles, the softening length 24 kpc$/h$.
Initial conditions are generated using {\tt mpgrafic}  \citep{mpgrafic}.
An Octree is built  to count  efficiently all particles 
within a given  sequence of concentric spheres of radii between $R=4,
5 \cdots $ up to $ 18 {\rm Mpc}/h $.  The center of these spheres is sampled regularly on a grid of 
$ 10\, {\rm Mpc}/h $ aside, leading to $50^{3}=125,000$ estimates of the density per snapshot.
Note that the cells overlap for radii larger than $5 \, {\rm Mpc}/h $.

\subsection{Density bias function}
\label{sec:measure-density-bias}
The density bias is estimated from spheres of radius $R=10 {\rm Mpc}/h $ that are separated by $r_{e}=20$Mpc$/h$ using the cross-correlations defined in equation~(\ref{eq:brho-cross}). More precisely, we compute a sum over each sphere $\rm I$ with density $\rho_{\rm I}$ and its 6 neighbours at distance $r_{e}=20$Mpc$/h$ labelled with the indices $\alpha_{{\rm I},j}$ for $1\leq j\leq 6$
\begin{equation}
\hat b (\hrho)=\frac 1 {\hat \xi}\!\left[\frac{\sum_{\rm I}\sum_{j=1}^{6}{\cal B}(\hrho\!-\!\Delta\rho/2\leq\rho_{\rm I}\leq\hrho\!+\!\Delta\rho/2)\rho_{\alpha_{{\rm I},j}}
}{
6\sum_{\rm I}{\cal B}(\hrho-\Delta\rho/2\leq\rho_{\rm I}\leq\hrho+\Delta\rho/2)}-1\right]\nonumber
\end{equation}
where ${\cal B}$ is a boolean function which evaluates to one if the density is in a bin centred on $\hrho$ with width $\Delta \rho = 3/21$ and the measured dark matter correlation function at distance $r_{e}$ is given by
\begin{equation}
\label{eq:xi}
\hat \xi (r_{e})=\frac{\sum_{\rm I=1}^{\rm N_{t}}\sum_{j=1}^{6}\rho_{\rm I}\rho_{\alpha_{{\rm I},j}}}{6\mathrm{N_{t}}}-1.
\end{equation}
In practice, we count all pairs of spheres only once when computing $\hat \xi (r_{e})$ by only considering three neighbours for each sphere.
{\red Error bars are then evaluated by computing the error on the mean of the density bias when the simulation is divided into eight sub-cubes.}

The density bias $\hat b(\hrho)$ measured for different values of the variance
 is shown
in the left-hand panel of Fig.~\ref{bias-rho-th}
and successfully compared to the predictions obtained in Section~\ref{sec:prediction}. 
In particular and as expected, it is found that the bias is null for densities of $1$. For count-in-cells, this trivially implies that imposing that the density equals its mean value within some sphere does not impact its neighbourhood sufficiently far enough.
The right-hand panel of Fig.~\ref{bias-rho-th} displays the corresponding residuals and confirms the extremely good agreement between theory and measurements.
Note that errors are relatively large for $\hrho \approx 1$ as $b(\hrho)$ is very close to zero in this region. Note also that the apparent discrepancy between theory and prediction for the lower density bin at $\sigma=0.39$ is mainly due to the fact that we did not take into account the size of the bin when predicting the density bias from equation~(\ref{eq:defbiasPDF}) while it would have been necessary especially in such steep regions of the plot. Eventually, we conclude that the prediction is within the error bars of the simulations for the whole range of densities probed by our simulation meaning that a 10\% accuracy at worst can be achieved with large-deviation theory. It is striking to see that the accuracy is poorer at higher redshift which could indicate that the main source of uncertainty here is not due to the theoretical prediction but to some numerical artefacts of our simulation. The (analytical) low-density approximation is shown to give a reasonable fit from $\approx 10\%$ precision for $\rho\lesssim 1$ to order one for larger densities. 

\begin{figure}
\includegraphics[width=\columnwidth]{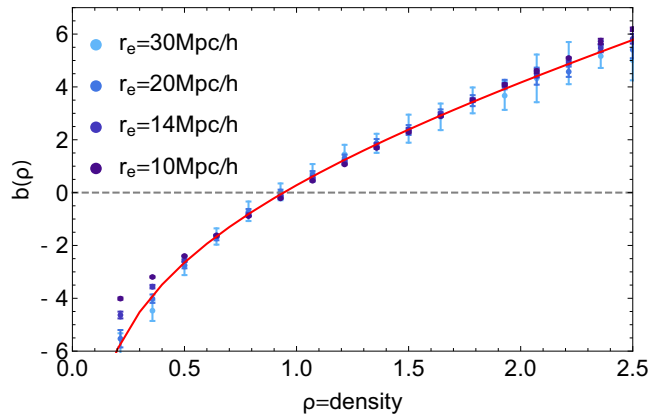}
   \caption{The density bias function $b(\rho)$ for $\sigma=0.97$ measured in our simulation for different separations $r_{e}$ as labelled. The large-separation prediction is displayed in red. 
   \label{bias-density}}
\end{figure}

In order to investigate the convergence towards the large-separation limit, we also measure the density bias at different separations $r_{e}$ (see Fig.~\ref{bias-density}, right-hand panel). For separation larger than $2R_{1}=20$Mpc$/h$ (adjacent spheres!), the prediction lies within the error bars of the simulations for the full range of density probed ($0.2<\hrho<2.5$). 

Note that appendix~\ref{sec:consistency} shows that the density bias estimated from the auto-correlation (see equation~\ref{eq:brho-auto}) is fully consistent with the cross-correlation based estimator used in this section.

\subsection{Slope bias function}

In order to measure the slope bias in the simulation, we consider the set of concentric spheres with radii $R_{1}=10 {\rm Mpc}/h $ and $R_{2}=11 {\rm Mpc}/h $ and pairs
that are separated by $r_{e}=20$Mpc$/h$ so that each set $\rm I$ of two concentric spheres with density $\rho_{\rm I}$ and slope $s_{\rm I}$ has six neighbours at distance $r_{e}$.
Following equation~(\ref{eq:srho-cross}), we compute again a sum over each set $\rm I$ and its 6 neighbouring sets at distance $r_{e}=20$Mpc$/h$ labelled with the indices $\alpha_{{\rm I},j}$ for $1\leq j\leq 6$
\begin{equation}
\hat b (\hs)\!=\!\frac 1 {\hat \xi}\!\left[\!\frac{\sum_{\rm I,j}{\cal B}(\hs\!-\!\Delta s/2\leq s_{\rm I}\leq \hs\!+\!\Delta s/2)\rho_{\alpha_{{\rm I},j}}
}{
6\sum_{\rm I}{\cal B}(\hs-\Delta s/2\leq s_{\rm I }\leq \hs+\Delta s/2)}\!-\!1\right]\!, \label{eq:effbs}
\end{equation}
where the bin width is set to $\Delta s=5/21$ here.

\begin{figure}
\includegraphics[width=\columnwidth]{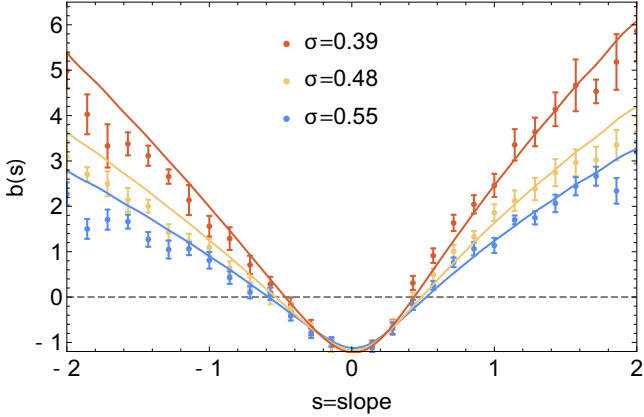}
   \caption{The slope bias $b(\hs)$ for different values of the variance as labelled. Measurements in the simulation for spheres of radii $R_{1}=10 {\rm Mpc}/h $ and $R_{2}=11 {\rm Mpc}/h $ separated by $r_{e}=20$Mpc$/h$ using the estimator defined in equation~(\ref{eq:effbs}) are displayed with error bars and compared with the prediction (solid line).  
   \label{sbias}}
\end{figure}

The slope bias function $b(\hs)$ measured for different values of the variance by means of equation~(\ref{eq:effbs})
is shown
in Fig.~\ref{sbias} and can be compared to the large-deviation
prediction displayed with solid lines. The two uncorrelated configurations predicted by our model are recovered around $\hs\approx\pm0.5$. 
One  could make use of this striking feature in order to minimize   large-scale clustering while restricting  counts to such slopes according to equation~(\ref{eq:var2cell})\footnote{The correlation vanishes at leading order in our calculations, not taken into account proximity effects and higher order correlation functions.}.
Predictions and measurements are in good agreement for small slopes $|\hs|\lesssim 1$ and both show that small slope regions are less clustered than sharper environments. 
They start to depart from one another in the tails of the distribution. Note that already the  PDF of the slope shows some level of discrepancy in this regime, which can be due to our numerical integration, our choice of parametrization for the variance (which does not take into account the running of the spectral index) or some unforeseen artifacts from the simulation. Further work is necessary to understand exactly why those discrepancies arise. It is hoped that  larger simulations and more robust analytical approximations  \cite[following][]{uhlemann16} will resolve these issues.

\begin{figure}
\includegraphics[width=\columnwidth]{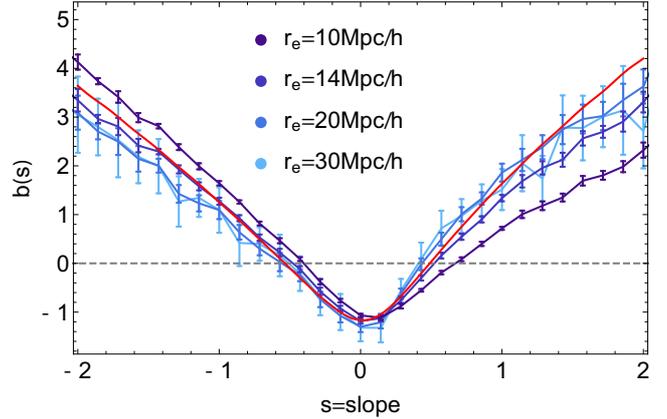}
   \caption{The slope bias  $b(\hs)$ measured in the simulation at redshift $z=0.97$ corresponding to $\sigma=0.48$ and for different values of the separation from $r_{e}=R_{1}=10$Mpc$/h$ to $3R_{1}=30 $Mpc$/h$ as labelled.  
   The measurements quickly converge towards the predicted large-separation limit (red solid line). \label{sbias-sep}}
\end{figure}

Fig.~\ref{sbias-sep} compares the measurements to the prediction for different values of the separation, $r_{e}=R_{1}$, $\sqrt2 R_{1}$ (along a diagonal), $2R_{1}$(adjacent cells) and $3R_{1}$.
The convergence towards the large-separation prediction (in red) is quick since the prediction starts to deviate from the error bars of the measurements only for $r_{e}\lesssim 2 R_{1}$ (adjacent cells!). Note that as expected, error bars are larger for larger separations.

\begin{figure}
\includegraphics[width=\columnwidth]{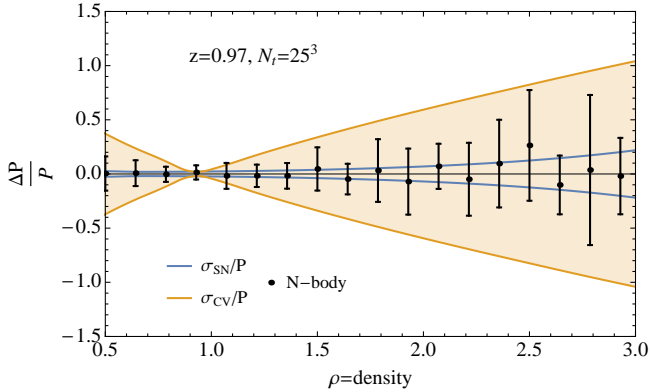}
   \caption{\red Expected shot noise (blue) and cosmic variance (yellow) for the density PDF measured from $25^{3}$ spheres of radius $R=10$Mpc$/h$ at redshift $z=1$ regularly drawn in our simulation. Those predictions are compared with error bars (black) estimated as the standard deviation among eight sub-cubes of the simulation. \label{fig:errors}}
\end{figure}

{\red
\section{Application to cosmic variance}
\label{sec:illustration}
We propose here to illustrate how the large-deviation prediction for the bias can be used to predict the statistics of errors. For that purpose, we measured the density PDF in $25^{3}$ spheres of radius $R=10$Mpc$/h$ at redshift $z=1$ in our simulation. As already pointed out in Section~\ref{sec:errPDF}, our estimate of the PDF can suffer from two types of errors : shot noise and cosmic variance.
We use equation~\ref{autocorr} to compute the expected error and display the result in Figure~\ref{fig:errors}. For the particular configuration used here, the cosmic variance is dominant compared to the shot noise error. As expected, error bars obtained by resampling seem to slightly underestimate the error budget. If the distance between the spheres increases, the amount of cosmic variance decreases as the spheres are less correlated. One can anticipate that for a given accessible volume, there is a balance to find between reducing the number of spheres to have them as independent as possible or increasing the number of spheres to reduce the shot noise. Finding the optimal number of spheres and radii to consider for a given survey geometry is left for future works.

}

\section{Conclusion}
%%%%%%%%%%
\label{sec:conclu}
We have shown how to compute joint statistics of the density within multiple concentric spheres in two regimes : 
i) the generating function of cumulants containing any powers of concentric densities in one location and one power of density at some \textit{arbitrary} distance
and ii) the two-point correlation function of the density in concentric spheres
when the cells are sufficiently well apart. The latter allowed us to
 estimate  the  bias and cosmic variance  involved  in applying standard concentric count-in-cells statistics
 to cosmological fields of finite extent in that regime. 
 The accuracy of the large-distance approximation was quantified against  numerical simulations and shown to be valid even for adjacent cells ($\approx 20$Mpc$/h$ here), strengthening earlier findings by \cite{1996A&A...312...11B} in the one-cell case. These simulations where also used to assess the validity  of the large-deviation principle, which only formally holds in the zero variance limit, but was shown to give accurate predictions even for variance of order unity, in the so-called quasi-linear regime. 
The bias functions ($b(\hrho)$, $b(\hs)$ in this work)  allow us to quantify  the covariances  expected in finite volume effects, hence build accurate maximum likelihood estimators   that could be applied to future surveys. In particular, the shape of $b(\hs)$ we found suggests that tailoring counts to cells which have slopes of the order of $\pm1/2$
  could be used to mitigate its effect. The formalism presented here can be straightforwardly extended to predict the bias of multiple concentric spheres including the bias of regions having a given density and slope $b(\hrho,\hs)$.

 While the large-deviation predictions of count-in-cell bias functions rely on  numerical integration in the complex plane, the recent results  of \cite{uhlemann16} sets the stage for accurate analytical approximations, using the logarithmic transform of the density.
 It would therefore be of  interest to apply this log-transformation to the two-point statistics presented in this paper. This will be the topic of upcoming work.
 
 {\red It has to be emphasized that the scale-dependence of the count-in-cell bias (that is clearly seen in simulations on scales below $\approx 20$ Mpc$/h$) can not be captured by the large-deviation principle used in this work. However, this formalism is a unique opportunity to get insights into the mildly non-linear evolution of halo biasing in contrast to peak models which usually assumes Gaussianity or very recently a Zel'dovich ballistic displacement of the initial peaks (\cite{2016MNRAS.456.3985B}, Baldauf et al., in prep). An hybrid analysis between both approaches is challenging but would be of great interest. Building up this peak theory in the large-deviation regime is left for future works.}
 
It would also be worth extending the investigation of biases and cosmic variance to the velocity field following the results obtained by \cite{1992ApJ...390L..61B} in the one-point case, and to projected densities.

\vspace{0.5cm}

{\bf Acknowledgements:}  
 This work is partially supported by the grants ANR-12-BS05-0002 and  ANR-13-BS05-0005 of the French {\sl Agence Nationale de la Recherche}.
The simulations were run on 
the {\tt Horizon} cluster. 
We acknowledge support from S.~Rouberol for running the cluster for us and warmly thank Cora Uhlemann for useful comments and 
careful reading of the manuscript. SC also thanks Marcelo Alvarez and Xin Wang for insightful discussions.

\bibliographystyle{mn2e}
\bibliography{LSStructure}

\appendix

\section{Radii decimations}
\label{decimation}

The purpose of this appendix is to insure that the expression of the cumulant generating function $\varphi(\{\lambda\})$, given by equation~(\ref{eq:phi2psi}),
is consistent with variable decimation, i.e.  we want to check that
\begin{multline}
\varphi(\{\lambda_{1},\dots,\lambda_{n}\})=
\\
 \varphi(\{\lambda_{1},\dots,\lambda_{n}, \lambda_{n+1}=0,\dots,\lambda_{n+m}=0\})\,,
\label{decimprop-m}
\end{multline}
where the left-hand side is computed from $n$ cells whereas the right-hand side is computed with $n+m$ cells. This property was also studied in BPC  but we propose here a slightly different proof and we will make use of some of the results -- in particular equation~(\ref{eq:taure}) -- in the main text.

By mathematical induction, it is sufficient to prove equation~(\ref{decimprop-m}) when $m$ is set to 1. We will therefore show here that the following property holds
\begin{equation}
\varphi(\{\lambda_{1},\dots,\lambda_{n}\})=
 \varphi(\{\lambda_{1},\dots,\lambda_{n}, \lambda_{n+1}=0\})\,,
\label{decimprop}
\end{equation}
where the left-hand side is still computed from $n$ cells whereas the right-hand side is now computed with $n+1$ cells.

Let us define a set $\mA$ of $n$ cells and a set $\mB$ of $m=1$ cell.
One can then define the covariance matrix  $\Sigma$ defined by $\Sigma_{ij}=\sigma^{2}(R_{i}\zeta(\tau_{i})^{1/3},R_{j}\zeta(\tau_{j})^{1/3})$ between two any cells of the union of $\mA$ and $\mB$. The covariance matrix restricted to $\mA$ will be denoted $\tilde \Sigma=\Sigma_{|\mA}$.  For the sake of clarity, in this appendix, greeks indices refer to the set $\mA$ and therefore go between 1 and $n$ and a tilde is displayed
 over the corresponding operators while in $\mA \cup\mB$ we use roman indices which go between 1 and $n+1$ and no tilde.

The cumulant generating functions for the $n$ cells in $\mA$ is given by
\begin{equation}
\varphi(\{\lambda_{1},\dots,\lambda_{n}\})=\lambda_{\mu}\tilde\rho_{\mu}-\frac{1}{2}\tilde \Xi_{\mu\nu}\tilde\tau_{\mu}\tilde\tau_{\nu}\,,
\end{equation}
with the stationary conditions 
\begin{equation}
\lambda_{\kappa}=\frac{\partial \tilde\Psi}{\partial\tilde \rho_k}=\tilde \Xi_{\mu\kappa}\tilde\tau_{\mu}\frac{\dd \tilde\tau_{\kappa}}{\dd\tilde \rho_{\kappa}}+
\frac{1}{2}\frac{\partial\tilde \Xi_{\mu\nu}}{\partial\tilde\rho_{\kappa}}\tilde\tau_{\mu}\tilde\tau_{\nu},
\label{lambdakappaexp}
\end{equation}
where $\tilde \Xi$ is the inverse matrix of $\tilde\Sigma$ and we use implicit summations over repeated indices $\mu,\nu$ between 1 and $n$.

Our purpose  is to show that it is identical to the expression of $\varphi(\{\lambda_{i}\}_{1\leq i\leq n+1})$ describing the cumulant 
generating function of the $n+1$ cells when the last value $\lambda_{n+1}$ is set to zero. In this latter case we have\begin{equation}
\varphi(\{\lambda_1,\dots,\lambda_{n},0\})=\lambda_{\mu}\rho_{\mu}-\frac{1}{2}\Xi_{ij}\tau_{i}\tau_{j}\,,
\end{equation}
with the stationary conditions 
\begin{eqnarray}
\lambda_{\kappa}&=&\Xi_{i\kappa}\tau_{i}\frac{\dd \tau_{\kappa}}{\dd \rho_{\kappa}}+
\frac{1}{2}\frac{\partial\Xi_{ij}}{\partial\rho_{\kappa}}\tau_{i}\tau_{j};\label{statAll1}\\
\lambda_{n+1}=0&=&\Xi_{n+1 i}\tau_{i}\frac{\dd \tau_{n+1}}{\dd \rho_{n+1}}+
\frac{1}{2}\frac{\partial\Xi_{ij}}{\partial\rho_{n+1}}\tau_{i}\tau_{j}\label{statAll2}\,,
\end{eqnarray}
with implicit summations over $i,j$ between 1 and $n+1$.
The constraint (\ref{statAll2}) allows one to determine the value of $\tau_{n+1}$
in terms of $\tau_{\nu}$.  
To do so, let us first express $\partial\Xi_{ij}/\partial\rho_{i}$ as a function of the derivatives of $\Sigma$. By differentiating $\Xi\cdot\Sigma=\mathbb{I}$, we get
\begin{equation}
\frac{\partial\Xi}{\partial\rho_{i}}=-\Xi\cdot\frac{\partial\Sigma}{\partial\rho_{i}}\cdot\Xi\,.
\end{equation}
One can also write this relation when the inverse matrix $\tilde \Xi$ is defined from the covariance matrix of the cells restricted in $\mA$ only, $\tilde\Sigma$,
\begin{equation}
\frac{\partial\tilde \Xi}{\partial\tilde \rho_{\kappa}}=-\tilde \Xi\cdot\frac{\partial\tilde\Sigma}{\partial\tilde \rho_{\kappa}}\cdot\tilde \Xi\,.
\end{equation}
Substituting $V=\Xi\cdot\tau$ and $\tilde V=\tilde \Xi\cdot\tilde \tau$ into equations~(\ref{lambdakappaexp}), (\ref{statAll1}), (\ref{statAll2}), we get
\begin{eqnarray}
\lambda_{\kappa}&=&\tilde V_{\kappa}\frac{\dd \tilde\tau_{\kappa}}{\dd \tilde\rho_{\kappa}}-
\frac{1}{2}\tilde V^{t}\cdot\frac{\partial\tilde\Sigma}{\partial\tilde\rho_{\kappa}}\cdot\tilde V,\\
\lambda_{\kappa}&=& V_{\kappa}\frac{\dd \tau_{\kappa}}{\dd \rho_{\kappa}}-
\frac{1}{2} V^{t}\cdot\frac{\partial\Sigma}{\partial\rho_{\kappa}}\cdot V,\\
0&=& V_{n+1}\frac{\dd \tau_{n+1}}{\dd \rho_{n+1}}-
\frac{1}{2}V^{t}\cdot\frac{\partial\Sigma}{\partial\rho_{n+1}}\cdot V\,.
\end{eqnarray}
An obvious  solution to this set of equations is given by 
\begin{equation}
V_{\mu}=\tilde V_{\mu}\,,\,V_{n+1}=0. \label{ap:solved}
\end{equation}
Indeed, in that case, using $\partial_{i}=\partial/\partial \rho_{i}$ and $\tilde\partial_{\kappa}=\partial/\partial \tilde \rho_{\kappa}$, we have $V_{\kappa}(\dd_{\kappa} \tau_{\kappa})=\tilde V_{\kappa}(\tilde\dd_{\kappa} \tilde\tau_{\kappa})$, $V^{t}\cdot\partial_{\kappa}\Sigma\cdot V=\tilde V^{t}\cdot\tilde\partial_{\kappa}\tilde\Sigma\cdot \tilde V+0$ and $V^{t}\cdot\partial_{n+1}\Sigma\cdot V=0$ because $\Sigma_{ij}$ solely depends on the densities $\rho_{i}$ and $\rho_{j}$.

The solution given by equation~(\ref{ap:solved}) is equivalent to
\begin{equation}
\label{eq:taure}
\tau_{\mu}=\tilde \tau_{\mu}\,,\,\tau_{n+1}=-\frac{\Xi_{\mu\, n+1}}{\Xi_{n+1\,n+1}}\tilde\tau_{\mu}=\sigma_{n+1\,\mu}\tilde \Xi_{\mu\nu}\tilde\tau_{\nu}\,.
\end{equation}
We then immediately get
\begin{equation}
\lambda_{\mu}\rho_{\mu}=\lambda_{\mu}\tilde\rho_{\mu}\,,\,
\frac 1 2 \tau_{i}V_{i}=
\frac 1 2 \tilde\tau_{\nu}\tilde V_{\nu}\,,
\end{equation}
so that the property given by equation~(\ref{decimprop}), namely $\varphi(\{\lambda_{1},\dots,\lambda_{n}\})=
 \varphi(\{\lambda_{1},\dots,\lambda_{n}, \lambda_{n+1}=0\})$, is established.
 Note that the same argument holds for the bias function so that $b_{\varphi}(\lambda_{1},\dots,\lambda_{n})=b_{\varphi}(\lambda_{1},\dots,\lambda_{n},0)$.
 
\section{Large distance limit }
\label{pkbgarg}

Let us present  an heuristic demonstration of the property (\ref{eq:cummixed}). The argument  is based on a peak-background type
construction. So let us consider a functional $\mF_{i}[\rho(\vx);\vx_{1}]$ of the density field $\rho(\vx)$ in the vicinity of the location $\vx_{1}$. The idea is 
that $\mF_{i}[\rho(\vx);\vx_{1}]$ depends on $\rho(\vx)$ only when $\vert\vx-\vx_{1}\vert$ is small enough, say $\vert\vx-\vx_{1}\vert\le \mR$
and we are interested in how
$\mF_{i}[\rho(\vx);\vx_{1}]$ and $\mF_{i}[\rho(\vx);\vx_{2}]$ are correlated when $\vert \vx_{1}-\vx_{2}\vert\gg R$. 
 One can then  say that the density field values at large separation are correlated only through the large scale fluctuations, or more precisely through the response function of the functional $\mF$ to large scale density fluctuations. We assume the latter to be in the linear regime
and we note them $\delta_{L}(\vx)$. Note that $\rho(\vx)$ is a non trivial functional of $\delta_{L}(\vx)$ through mode coupling.
Formally in the vicinity of $\vx_{1}$ for instance we could write
\begin{equation}
\rho(\vx)=\rho_{S}(\vx)+\mD_{\mL}[\rho(\vx);\vx_{1}]\delta_{\mL}(\vx_{1})+\dots.
\end{equation}
where $\rho_{S}(\vx)$ is the value of the field when long wave modes have been suppressed
and where $\mD_{\mL}[\rho(\vx);\vx_{1}]$ gives the linear response function of the field $\rho(\vx)$ to the large scale fluctuation $\delta_{\mL}$
in the vicinity of $\vx_{1}$.
Then any functional $\mF[\rho(\vx);\vx_{1}]$ of the density field can be similarly expanded as a function of the local density contrast
\begin{equation}
\mF[\rho(\vx);\vx_{1}]=\mF[\rho_{S}(\vx);\vx_{1}]+\mD_{\mL}[\mF[\rho(\vx)];\vx_{1}]\delta_{\mL}(\vx_{1})+\dots.
\end{equation}
Note that in this expression both $\mF[\rho_{S}(\vx);\vx_{1}]$ and $\mD_{\mL}[\mF[\rho(\vx)];\vx_{1}]$
are random quantities that depend on the small scale density fluctuations, independently on the long wave modes
collected in $\delta_{\mL}(\vx_{1})$.
When taking ensemble average of combination of such quantities,  one can observe that
\begin{equation}
\langle \mF[\rho(\vx);\vx_{1}]\rangle = \langle \mF[\rho_{S}(\vx)]\rangle
\end{equation}
at dominant order in $\delta_{\mL}$ and then that
\begin{eqnarray}
\langle \mF[\rho(\vx);\vx_{1}]\ \rho(\vx_{2})\rangle & = &\langle \mF[\rho_{S}(\vx);\vx_{1}]\rangle\nonumber 
\\&&\hspace{-.8cm}+\ 
\langle \mD_{\mL}[\mF[\rho_{S}(\vx)];\vx_{1}] \rangle\ \xi_{\mL}(\vx_{1},\vx_{2})\nonumber 
\\&&\hspace{-.8cm}+\ \dots
\end{eqnarray}
where
\begin{equation}
\xi_{\mL}(\vx_{1},\vx_{2})\equiv \langle \delta_{\mL}(\vx_{1})\ \delta_{\mL}(\vx_{2})\rangle.
\end{equation}
The key point is to further observe that,
\begin{eqnarray}
\langle \mF[\rho(\vx);\vx_{1}]\ \mF[\rho(\vx);\vx_{2}] \rangle & = &\nonumber\\
&&\hspace{-3.9cm}\langle \mF[\rho_{S}(\vx);\vx_{1}]\rangle
\langle \mF[\rho_{S}(\vx);\vx_{2}]\rangle
\nonumber \\
&&\hspace{-3.9cm}+
\langle\mD_{\mL}[\mF[\rho_{S}(\vx)];\vx_{1}] \rangle\ 
\langle \mD_{\mL}[\mF[\rho_{S}(\vx)];\vx_{2}] \rangle\ 
\xi_{\mL}(\vx_{1},\vx_{2})
\nonumber \\
&&\hspace{-3.9cm}+\ \dots
\label{oneptproprelation}
\end{eqnarray}
When taking the connected part of these moments, these forms lead to the functional relation,
\begin{eqnarray}
\langle \mF[\rho(\vx);\vx_{1}]\ \mF[\rho(\vx);\vx_{2}] \rangle_{c}&=&\frac{1}{\xi_{\mL}(\vx_{1},\vx_{2})}
\nonumber \\
&&\hspace{-2.9cm}\times
\langle \mF[\rho(\vx);\vx_{1}]\ \rho(\vx_{2})\rangle_{c} \langle \mF[\rho(\vx);\vx_{2}]\ \rho(\vx_{1})\rangle_{c}.
\end{eqnarray}
Note that it can easily be extended  to two different functionals of the density field and 
when applied to products of density fields, it precisely leads to the relation~(\ref{eq:cummixed}).
This construction actually reproduces the expansion scheme developed in \cite{2008PhRvD..78j3521B} in Fourier space.
In this context the $\langle \mD_{\mL}[\mF[\rho_{S}(\vx)];\vx_{1}] \rangle$ is simply the nonlinear propagator in real 
space. The relation~(\ref{oneptproprelation}) could be extended to higher order contributions provided one knew how to compute 
multi-point propagators.

\section{Cosmic error statistics}
\label{discretecounts}

In this section, we show how to go from equation~(\ref{eq:fullPDFlargeseparation}) to the cosmic variance of discrete counts.
Hence let us assume that the joint PDF of the density in $\rm N_{t}$ cells reads
\begin{equation}
\mP(\{\hrho_1\},\dots,\{\hrho_{\rm N_{t}}\}) =\prod_{\rm I=1}^{\rm N_{t}} \mP(\hrho_{\rm I})
\left[1+\
\sum_\mathrm{\rm i<j}
b(\hrho_{\rm I}) b(\hrho_{\rm J}) \xi_{\rm IJ}
\right]\,.
\label{eq:defbiasPDF-C}
\end{equation}
In what follows, we will take $\xi_{\rm IJ}=\xi$ independent from the positions and we will denote the density bias $b_{\rm I}=b(\hrho_{\rm I})$.

\subsection{The one-cell count-in-cell PDF}
We want to estimate $\mP(\hat\rho)\dd \hat\rho$ by measuring the density in ${\rm N_{t}}$ spheres of same radius $R$.
If the spheres are independent i.e there is no bias in the measure $b_{\rm I}=0$, then the probability of having N spheres with density $\hat\rho\pm\Delta\rho/2$ is simply given by a binomial distribution
\begin{equation}
P^{\rm {unbiased}}({\rm N})=\left(\!\begin{array}{c}
 {\rm N_{t}} \\
{\rm N} \\ 
\end{array} \!\right)p^{\rm N}(1-p)^{\rm N_{t}-N}\,,
\end{equation}
with $\left(\!\begin{array}{c}
\rm  N_{t} \\
\rm N \\ 
\end{array} \!\right)=\rm N_{t}!/N!/(N_{t}-N)!$ and $p$, the probability for the density of one sphere to be in $\hat \Delta=[\hat\rho-\Delta \rho/2,\hat\rho+\Delta\rho/2]$, is given by
\begin{equation}
p=\int_{\hat\Delta}\dd\rho\,\mP(\rho)\,.
\end{equation}
In the Poisson limit (where $\rm N_{t}\gg N$ and $p\rm N_{t}$ is constant), using the Stirling formula $x!\approx \sqrt{2\pi x} \left({x}/e\right)^{x}$ when $x$ goes to infinity, this probability distribution trivially becomes a Poisson distribution parametrized by ${\rm \bar N}=p \rm N_{t}$
\begin{equation}
P^{\rm{unbiased}}(\rm N)\approx\rm \frac{\bar N^{N}}{N!}\exp(-\bar N)\,.
\end{equation}

Accounting for a non-zero bias in equation~\ref{eq:defbiasPDF-C}, the probability of having N spheres with density in
 $\hat \Delta=\hat\rho\pm\Delta\rho/2$ becomes
\begin{equation}
P^{\rm{biased}}({\rm N})\!=\!\!\left(\!\!\!\begin{array}{c}
\rm N_{t} \\
\rm N \\ 
\end{array} \!\!\!\!\right)
P\!\!
\left(\!\!\!\begin{array}{c}
\hrho_{1},\!\dots\!,\hrho_{\rm N}\in \hat\Delta \\
\hrho_{\rm N+1},\!\dots\!, \hrho_{\rm N_{t}} \notin\hat \Delta  \\ 
\end{array} \!\!\!\!\right),
\label{eq:Pbiased}
\end{equation}
where the probability of having only the first N densities in $\hat \Delta$ reads
\begin{multline}
P\!\!
\left(\!\!\!\begin{array}{c}
\hrho_{1},\!\dots\!,\hrho_{\rm N}\in \hat\Delta \\
\hrho_{\rm N+1},\!\dots\!, \hrho_{\rm N_{t}} \notin\hat \Delta  \\ 
\end{array} \!\!\!\!\right)
 =\\
\left(\int_{\hat\Delta}\right)^{\!\!\!\rm N}\!\!\dd\hrho_{1}\!\cdots\!\dd\hrho_{\rm N}\!\left(\!\int_{\mathbb{R}_{+}/\hat\Delta}\!\right)^{\!\!\!\rm N_{t}-N}\!\!\!\!\!\!\!\!\!\!\!\!\dd\hrho_{\rm N+1}\!\cdots\!\dd\hrho_{\rm N_{t}}
%\\\times
\mP(\hrho_1,\dots,\hrho_{\rm N_{t}}).
\nonumber
\end{multline}
Given equation~(\ref{eq:defbiasPDF-C}), equation~(\ref{eq:Pbiased}) eventually yields
\begin{align}
P^{\textrm{biased}}(\rm N)&=\left(\!\begin{array}{c}
 \rm N_{t} \\
\rm N \\ 
\end{array} \!\right)\times\left[p^{\rm N}(1-p)^{\rm N_{t}-N}\right.\nonumber\\
+&\xi \frac{\rm N(N-1)}{2}(pb)^{2}p^{\rm N-2}(1-p)^{\rm N_{t}-N}\nonumber\\
+&\xi {\rm N (N_{t}-N)} p^{\rm N-1} pb (-pb)(1-p)^{\rm N_{t}-N-1}\nonumber\\
+&\xi \frac{\rm (N_{t}-N)(N_{t}-N-1)}{2}p^{\rm N}(-pb)^{2}\left.(1-p)^{\rm N_{t}-N-2}
\right]\nonumber,
\end{align}
where we introduced $pb$ such that
\begin{equation}
pb=\int_{\hat \Delta}\dd\hrho\,\mP(\hrho)b(\hrho)
\end{equation}
that comes with complement $-pb$ as the normalisation of $\mP$ enforces $\int_{\mathbb{R}_{+}}\dd\hrho\,\mP(\hrho)b(\hrho)=0$.
In full generality, note that the factors $\xi$ entering in the expression of the biased count probability should be understood respectively (and in the order of appearance) as the mean i) autocorrelation between spheres of density in $\hat \delta$; ii) cross-correlation between spheres of that density and the others; iii) auto-correlation between the spheres of density not in $\hat \delta$.
For the sake of simplicity, we will consider here that it is given by the mean correlation $\xi={2}\sum_{\rm I<J}\xi_{\rm IJ}/{[\rm N_{t}(N_{t}-1)]}$.

The Poisson limit is then simply given by
\begin{align}
P^{\textrm{biased}}({\rm N})\approx \frac{\rm\bar N^{\rm N}}{\rm N!}e^{\rm-\bar N}\!\!\left[1+b^{2}\xi \!\left(\rm\frac{N(N-1)}{2}-N\bar N +\frac{\bar N^{2}}{2}\right)\right]\,,\nonumber
\end{align}
where $b=pb/p$.
It is then straightforward to check that the probability distribution $P^{\textrm{biased}}$ 
i) is normalised;
ii) has mean $\left\langle\rm N\right\rangle=\rm \bar N$ (and is therefore unbiased);
iii) has variance $\left\langle {\rm N}^{2}\right\rangle-\left\langle {\rm N}\right\rangle^{2}=\bar {\rm N}+b^{2}\xi\bar {\rm N^{2}}$. For large enough $\rm \bar N$, sampling errors can be neglected and the cosmic variance is directly proportional to $b^{2}$.

\subsection{Cross-correlations}
Let us now show how the estimate of the one-cell PDF in two distinct bins are correlated.
If the spheres are independent (i.e the joint PDF is unbiased), then the probability, $P^{\rm{unbiased}}(\rm N_{1},N_{2})$, of getting $\rm N_{1}$ spheres with density in $\hat\Delta_{1}$ and $\rm N_{2}$ spheres with density in $\hat\Delta_{2}$ is given by
\begin{multline}
\hskip -0.4cm P^{\textrm {unbiased}}\!\!=
\left(\!\!\begin{array}{c}
\! \rm N_{t}\! \\
\!\rm N_{1}\! \\ 
\end{array} \!\!\right)\!
\left(\!\!\begin{array}{c}
\!\rm  N_{t}-N_{1}\! \\
\! \rm N_{2}\! \\ 
\end{array} \!\!\right)
p_{1}^{\rm N_{1}}p_{2}^{\rm N_{2}}(1-p_{1}-p_{2})^{\rm N_{t}-N_{1}-N_{2}}\,,\nonumber
\end{multline}
where $p_{1}$ is the probability for the density of one sphere to be in $\hat\Delta_{1}$ and $p_{2}$ to be in $\hat\Delta_{2}$.
In the Poisson limit, this probability distribution trivially becomes the product of two Poisson distribution parametrized by ${\rm \bar N_{1}}=p_{1}\rm  N_{t}$ and ${\rm \bar N_{2}}=p_{2}\rm N_{t}$
\begin{equation}
P^{\textrm{unbiased}}({\rm N_{1},N_{2}})\approx P^{\textrm{unbiased}}({\rm N_{1}}) P^{\textrm{unbiased}}({\rm N_{2}}) \,.
\end{equation}
As expected, in the case where there is no spatial correlation between spheres, the estimate of the density PDF in each bin is independent.

In the biased case, the probability of having $\rm N_{1}$ spheres with density in $\hat\Delta_{1}$ and $\rm N_{2}$ spheres with density in $\hat\Delta_{2}$ becomes
\begin{equation}
P^{\textrm{biased}}(\rm {N_{1},N_{2}})=P^{\textrm {unbiased}}({\rm N_{1},N_{2}})\left[1+\xi \mathfrak{b}\right]\,, \label{eq:PN12}
\end{equation}
where $\mathfrak{b}$ is given by
\begin{align}
\mathfrak{b}&=\frac{\rm N_{1}(N_{1}-1)}{2}b_{1}^{2}+{\rm N_{1}N_{2}}b_{1}b_{2}+\frac{\rm N_{2}(N_{2}-1)}{2}b_{2}^{2}\nonumber\\
&+\frac{\rm (N_{t}-N_{1}-N_{2})(N_{t}-N_{1}-N_{2}-1)}{2}\left(\frac{b_{1}p_{1}+b_{2}p_{2}}{1-p_{1}-p_{2}}\right)^{2}\nonumber\\
&+{\rm (N_{1}}b_{1}+{\rm N_{2}}b_{2}){\rm N_{t}}\frac{b_{1}p_{1}+b_{2}p_{2}}{1-p_{1}-p_{2}}\,\nonumber,
\end{align}
the bias parameters $b_{1}$ and $b_{2}$ being defined as
\begin{equation}
b_{i}=\frac{\int_{\hat\Delta_{i}}\dd\hrho\,\mP(\hrho)b(\hrho)}{\int_{\hat\Delta_{i}}\dd\hrho\,\mP(\hrho)}\,.
\end{equation}
In the Poisson limit, equation~(\ref{eq:PN12}) becomes
\begin{align}
P^{\textrm{biased}}({\rm N_{1},N_{2}})&\approx P^{\textrm {unbiased}}(\rm N_{1},N_{2})\left[\frac{}{}1\right.\nonumber\\
&+\xi b_{1}^{2}\left(\frac{\rm N_{1}(N_{1}-1)}{2}+\frac{\rm \bar N_{1}^{2}}{2}-\rm N_{1}\bar N_{1}\right)\nonumber\\
&+\xi b_{2}^{2}\left(\frac{\rm N_{2}(N_{2}-1)}{2}+\frac{\rm \bar N_{2}^{2}}{2}-\rm N_{2}\bar N_{2}\right)\nonumber\\
&+\xi b_{1}b_{2}\left.\left(\rm N_{1}N_{2}+\bar N_{1}\bar N_{2}-N_{1}\bar N_{2}-\bar N_{1}N_{2}\right)\frac{}{}\right]\nonumber.
\end{align}
This probability distribution $P^{\textrm{biased}}(\rm N_{1},N_{2})$ is normalised, its marginals are respectively $P^{\textrm{biased}}(\rm N_{1})$ and $P^{\textrm{biased}}(\rm N_{2})$ and the covariance between $\rm N_{1}$ and $\rm N_{2}$ is given by
\begin{equation}
\left\langle{\rm N_{1}N_{2}} \right\rangle={\rm\bar N_{1}\bar N_{2}}(1+\xi b_{1}b_{2}).
\end{equation} 

It should be straightforward, if tedious, to generalize equation~(\ref{eq:PN12}) to an arbitrary number of cells. 
\section{Consistency of bias}
\label{sec:consistency}

Let us measure $b^{2}(\hrho)$ from two estimators and check for consistency.
The first  is based on cross-correlations and used in Section~\ref{sec:measure-density-bias} to measure the density bias function (see equation~(\ref{eq:brho-cross})). We use the following estimator to measure $b^{2}(\hrho)$ in this context,  
\begin{equation}
\label{eq:b2rho-cross}
\hat b^{2}_{\rm cross} (\hrho)=\frac 1 {\hat \xi^{2}}\!\left[\frac{\sum_{\rm I}\sum_{j=1}^{6}\epsilon(\rho_{\rm I},\hrho,\hat\Delta)
\rho_{\alpha_{{\rm I},j}}
}{
6\sum_{\rm I}
\epsilon(\rho_{\rm I},\hrho,\hat\Delta)
}-1\right]^{2},
\end{equation}
where the sum is over each sphere $1\leq \rm I \leq N_{t}$ of radius $R=10 {\rm Mpc}/h $ with density $\rho_{\rm I}$ and its 6 neighbours at distance $r_{e}=20$Mpc$/h$ labelled with the indices $\alpha_{{\rm I},j}$ for $1\leq j\leq 6$, $\epsilon(x,\hrho,\hat\Delta)={\cal B}(x\in[\hrho\!-\!\Delta\rho/2,\hrho\!+\!\Delta\rho/2]$
with ${\cal B}$ a boolean function which evaluates to one if the density is in a bin centred on $\hrho$ with width $\Delta \rho = 3/21$ and the measured dark matter correlation function at distance $r_{e}$ is given by equation~(\ref{eq:xi}).

Alternatively, one can use equation~(\ref{eq:brho-auto}) to estimate the density bias squared via the auto-correlation of cells
\begin{equation}
\label{eq:b2rho-auto}
\hat b^{2}_{\rm auto} (\hrho)\!=\!\frac 1 {\hat \xi}\!\!\left[\!\frac{{\rm N_{t}}\sum_{{\rm I},j}
\!\epsilon(\rho_{\rm I},\hrho,\hat\Delta)
\epsilon(\rho_{\alpha_{{\rm I},j}},\hrho,\hat \Delta)
}{
6\sum_{\rm I}
\epsilon(\rho_{\rm I},\hrho,\hat\Delta)
\sum_{{\rm I},j}\epsilon(\rho_{\alpha_{{\rm I},j}},\hrho,\hat\Delta)
}\!-\!1\!\right]\!\!.
\end{equation}

\begin{figure}
\includegraphics[width=\columnwidth]{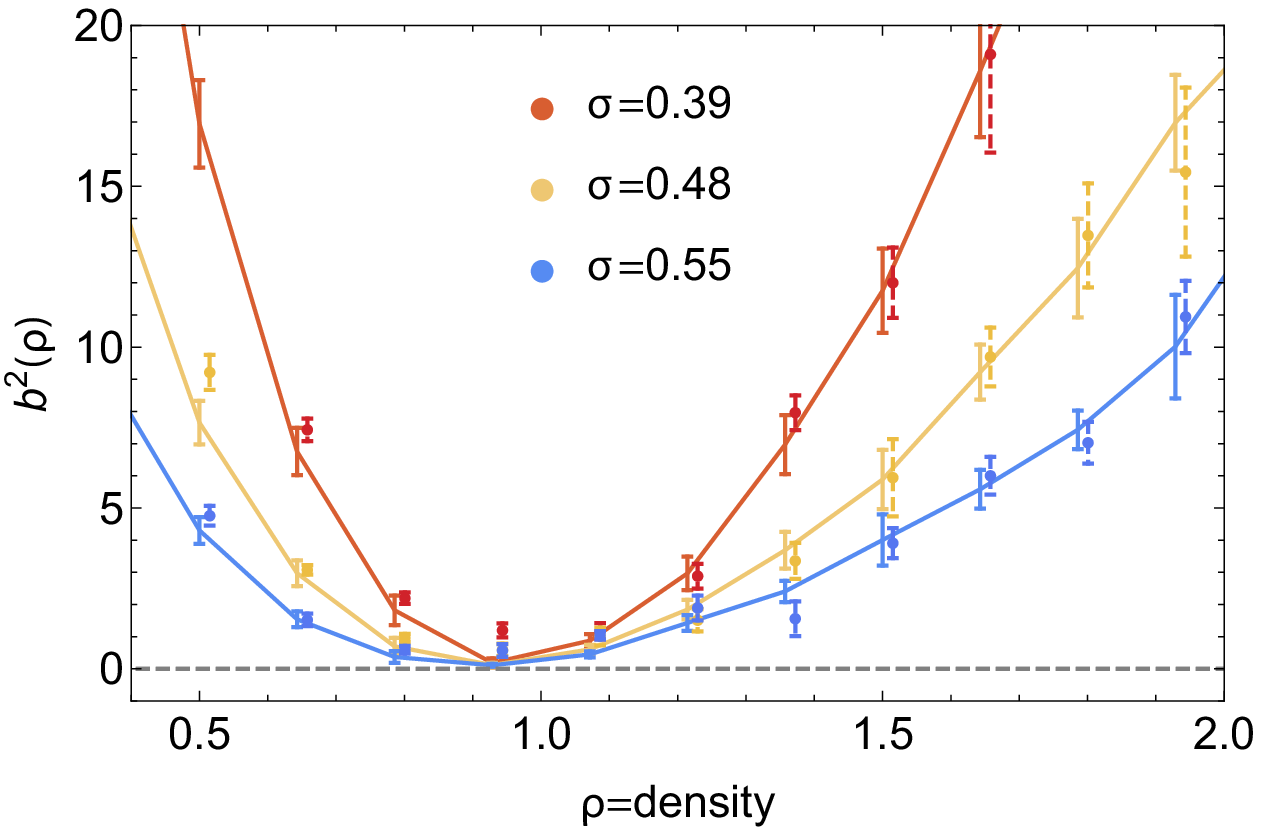}
   \caption{Density bias squared measured in the simulation from the auto-correlation (dashed error bars) and cross-correlation (solid line and error bars) estimators given by equations~(\ref{eq:b2rho-cross}) and (\ref{eq:b2rho-auto}) at different variances as labelled. Both approaches give consistent results.    \label{fig:comparison-rhobias}}
\end{figure}

Fig.~\ref{fig:comparison-rhobias} compares the mean and error on the mean of the density bias estimated by equations~(\ref{eq:b2rho-cross}) and (\ref{eq:b2rho-auto}) when the simulation is divided into eight sub-cubes. Both approaches seem to give consistent result in particular for intermediate densities. Note that in the case of the auto-correlation (equation~(\ref{eq:b2rho-auto})), the statistics is expected to be poorer and the estimator therefore noisier. Indeed, the frequency of occurrence of the event ``the two densities are in the same bin'' is low in particular in the tails of the distribution and it is very likely that our error bars are under-estimated in this case.

\section{Asymptotic behaviour}
\label{sec:largedensity}

The asymptotic behaviour of the density bias $b(\hrho)$ can be predicted by integrating equations~(\ref{eq:PDF}) and (\ref{eq:defbiasPDF}) around the critical point which dominates the complex integration for $\hrho$ sufficiently large compared to $\rho_{c}$.
\cite{Bernardeau15} indeed showed that in this regime, the cumulant generating function behaves like
\begin{equation}
\label{eq:varphiasympt}
\varphi(\lambda)=\varphi_{c}+(\lambda-\lambda_{c})\rho_{c}+(\lambda-\lambda_{c})^{3/2}\frac{2\sqrt2}{3\sqrt{\pi_{3}}}-(\lambda-\lambda_{c})^{2}\frac{\pi_{4}}{6\pi_{3}^{2}}+\cdots\nonumber
\end{equation}
with $\pi_{i}=\Psi^{(i)}(\rho_{c})$, the successive derivatives of the rate function at the critical point.
This expansion for the cumulant generating function can be obtained by first inverting, around the critical point,  the relation between $\rho$ and $\lambda$ coming from equation~(\ref{stationary1})
\begin{equation}
\lambda=\lambda_{c}+\sum_{i}\frac{\left(\rho-\rho_{c}\right)^{i}}{i!}\pi_{i+1}\,,
\end{equation}
which reads
\begin{equation}
 \rho=\rho_{c}
 +\sqrt{\frac{2(\lambda-\lambda_{c})}{\pi _3}}
 -\frac{\pi _4 (\lambda-\lambda_{c} )}{3 \pi _3^2}
   +
   \cdots,\nonumber
   \end{equation}
and plugging the result into equation~(\ref{eq:phi2psi}).

\begin{figure}
\begin{center}
\includegraphics[width=0.8\columnwidth]{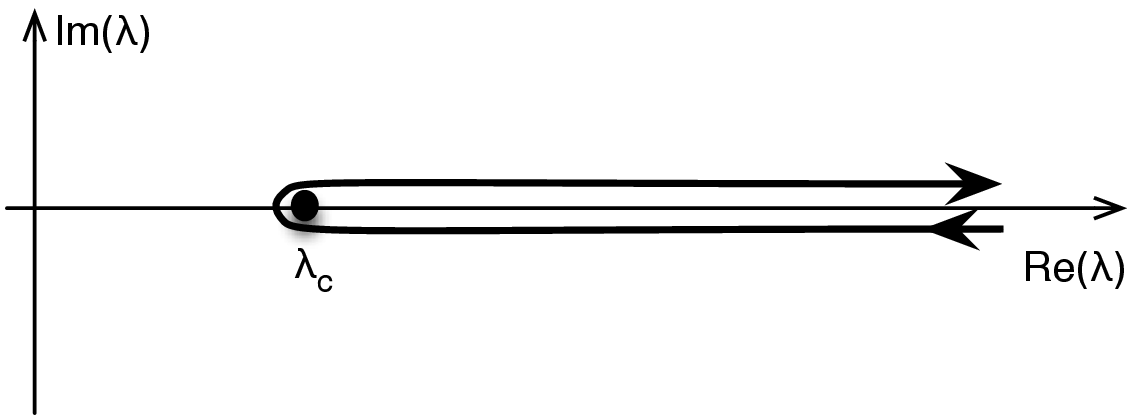}
   \caption{The path line in the $\lambda$ complex plane used for the computation of the large-density asymptotic forms follows a so-called Hankel contour.
   \label{ContourAsymp}
   }
   \end{center}
\end{figure}

The large-density tail of the density PDF and density bias can then be derived by computing the
inverse Laplace transform of the generating function $\varphi(\lambda)$ when the integrand of  equations~(\ref{eq:PDF}) and (\ref{eq:defbiasPDF}) is dominated by its
singular part, near $\lambda\approx \lambda_{c}$. 
In that case, \cite{1989A&A...220....1B} and BPC showed that the dominant contribution of the path in the complex plane is along the real axis and wrapping around the singular value $\lambda_{c}$ as illustrated on Fig.~\ref{ContourAsymp}. As the two branches $\mathbb{R}_{+}+\lambda_{c}\pm\ii\epsilon$ of that path yield complex conjugate contributions, the density PDF can be approximated by
\begin{multline}
P(\hrho)\approx
\Im\!\left\{\!
\int_{\ii\epsilon+\lambda_{c}}^{\ii\epsilon+\infty}
\!\frac{\dd\lambda}{\pi}\!
\exp[{\varphi_{c}\!-\!\lambda_{c}\hrho\!-\!(\lambda\!-\!\lambda_{c})(\hrho\!-\!\rho_{c})}]\right.\\
\times \left.\left[1+\frac{2\sqrt2}{3\sqrt{\pi_{3}}}(\lambda\!-\!\lambda_{c})^{3/2}+\dots
\right]\right\}\,,
\label{eq:asymptPDF}
\end{multline}
and the density bias by
\begin{multline}
\!\!b(\hrho)P(\hrho)\approx
\Im\!\left\{\!
\int_{\ii\epsilon+\lambda_{c}}^{\ii\epsilon+\infty}
\!\frac{\dd\lambda}{\pi}\!
\exp[{\varphi_{c}\!-\!\lambda_{c}\hrho\!-\!(\lambda\!-\!\lambda_{c})(\hrho\!-\!\rho_{c})}]\right.\\
\times \left.\left[1+\frac{2\sqrt2}{3\sqrt{\pi_{3}}}(\lambda\!-\!\lambda_{c})^{3/2}+\dots
\right]b_{\varphi}(\lambda)\right\}\,,
\label{eq:asymptbias}
\end{multline}
where $\Im$ is the imaginary part. 

As already shown in \cite{Bernardeau15}, the resulting large-density PDF can then be written, to leading order, as
\begin{equation}
\mP(\hrho\gg\rho_{c})\!=\!\frac{1}{\pi}\exp\left(\varphi_{c}-\lambda_{c}\hrho\right)\Im\!\left[\frac{\sqrt{\pi}}{\sqrt{2\pi_{3}}}(\hrho-\rho_{c})^{-5/2}+\cdots\right]\nonumber\,,
\end{equation}
while the corresponding bias function is given to leading order as  the ratio of equations~(\ref{eq:asymptbias}) and (\ref{eq:asymptPDF})
\begin{multline}
b(\hrho\gg\rho_{c})=(\hrho-\rho_{c})
V^{(1)}(\rho_{c})\\ \hskip 0.25cm
\!+\!V(\rho_{c})\!-\!\frac{\pi_{4}}{2\pi_{3}^{2}}V^{(2)}(\rho_{c})\!+\!\frac{1}{2\pi_{3}}V^{(3)}(\rho_{c})\!+\!{\cal O}\left(\!\frac{1}{\hrho-\rho_{c}}\!\right)\,
\end{multline} 
where 
 in the one-cell case $V(\rho)=\Xi_{ij}\tau_{j}(\rho_{j}$ is given by $V(\rho)=\zeta^{-1}(\rho)/\sigma^{2}(R\rho^{1/3},R\rho^{1/3})$ and is nothing but the bias function $V(\rho)=b_{\varphi}(\lambda=\Psi'(\rho))$.
In particular, one can compute the successive derivatives of $V(\rho)$ at the critical density, for instance
\begin{equation}
V^{(1)}(\rho_{c})=\frac {(\lambda_{c}\rho_{c}(\rho_{c}^{1/\nu}-1-1/\nu)+\frac{1}{\nu}\varphi_{c})
}{
\nu(\rho_{c}^{1/\nu}-1)^{2}\rho_{c}^{1-1/\nu}}\,.
\end{equation}

\section{Approximation schemes for the slope}
\label{sec:approx-slope}
In the context of large-deviation theory, 
one can compute the rate function of the slope $s=(\rho_{2}-\rho_{1})R_{1}/\Delta R $
by applying the so-called contraction principle which yields
\begin{equation}
\Psi_{s}(\hs)=\inf_{\rho} \Psi\left(\rho,\rho+s\frac{\Delta R}{R_{1}}\right)\,.
\end{equation}
The cumulant generating function of the slope can then be obtained through the Gartner-Ellis theorem as the Legendre transform of the rate function 
\begin{equation}
\varphi_{s}(\mu)=\mu s -\Psi_{s}(s)\quad{ \textrm {with}}\quad \mu=\Psi_{s}'(s)\,.
\end{equation}
The slope PDF then follows by Inverse Laplace transform as shown in equation~(\ref{eq:defslopePDF})).
It is now easy to see that the approximations of the slope PDF and bias can be obtained similarly to what has been done for the density.
In this case, the slope cumulant generating function has typically two critical points $s_{c}^{-}<0$ and $s_{c}^{+}>0$ corresponding to the zeros of the second derivative of the rate function $\dd^{2}\Psi_{s}(s)/\dd s^{2}=0$.

For small slope $s_{c}^{-}<s<s_{c}^{+}$, the slope PDF and bias can be approximated by a saddle point approximation of equations~(\ref{eq:defbiasPDF-slope}) and (\ref{eq:defslopePDF})  so that
\begin{equation}
{\cal P}(\hs)\approx\sqrt{\frac{\Psi_{s}''(\hs)}{2\pi}}\exp\left(-\Psi_{s}(\hs)\right)\,,
\end{equation}
and
\begin{equation}
b(\hs)\approx b_{\varphi}(-R_{1}\Psi_{s}'(s)/\Delta R,R_{1}\Psi_{s}'(s)/\Delta R)\,.
\end{equation}
However, this approximation is not sufficient (in particular it would predict $b(0)=0$ which is far from the true value);  one should Taylor expand the bias function $b_{s}$ around $\mu=\Psi_{s}'(\hat s)$ in order to get a better fit to the full numerical integration. This issue is left for future work.

Conversely, the asymptotes of the PDF and bias at large (positive and negative) slopes can be obtained similarly to what was done for the large-density regime (see appendix~\ref{sec:largedensity}). For instance, for large positive slope, the PDF can be approximated by
\begin{multline}
\mP(\hs\gg s_{c}^{+})\!=\!\frac{1}{\pi}\exp\left(\varphi_{s}(\mu_{c}^{+})-\mu_{c}^{+}\hs\right)\\
\times
\Im\!\left[\sqrt{\frac{{\pi}}{{2\pi^{s,+}_{3}}}}(\hs-s_{c}^{+})^{-5/2}]\right.\left.
+{\cal O}\left((\hs-s_{c}^{+})^{-7/2}\right)\right]\nonumber\,,
\end{multline}
with $\pi_{i}^{s+}=\Psi_{s}^{(i)}(s_{c}^{+})$, 
while the corresponding slope bias is
\begin{multline}
b(\hs\gg s_{c}^{+})=(\hs-s_{c}^{+})
V_{s}^{(1)}(s_{c}^{+})
\!+\!V_{s}(s_{c}^{+})\!-\!\frac{\pi^{s}_{4}}{2(\pi^{s}_{3})^{2}}V_{s}^{(2)}(s_{c}^{+})\!\\
+\!\frac{1}{2\pi^{s}_{3}}V_{s}^{(3)}(s_{c}^{+})\!+\!{\cal O}\left(\!\frac{1}{\hs-s_{c}^{+}}\!\right)\,,\nonumber
\end{multline} 
where we define  $V_{s}(s)=b_{\varphi}(-R_{1}\Psi_{s}'(s)/\Delta R,R_{1}\Psi_{s}'(s)/\Delta R)$.

\end{document}